\documentclass[fleqn,usenatbib]{mnras}

\usepackage{newtxtext,newtxmath}

\usepackage[T1]{fontenc}

\DeclareRobustCommand{\VAN}[3]{#2}
\let\VANthebibliography\thebibliography
\def\thebibliography{\DeclareRobustCommand{\VAN}[3]{##3}\VANthebibliography}

\usepackage{graphicx}	
\usepackage{amsmath}	

\usepackage{amssymb}	
\usepackage{soul}
\usepackage{color}

\definecolor{mygreen}{rgb}{0.19,0.55,0.11}

\title[Linking LMXBs, MSPs, UCXBs and GW sources]{Formation of millisecond pulsars with helium white dwarfs, ultra-compact X-ray binaries and gravitational wave sources}

\author[Chen et al.]{
Hai-Liang Chen,$^{1,2,3}$\thanks{E-mail: chenhl@ynao.ac.cn}
Thomas M. Tauris,$^{3,4}$
Zhanwen Han,$^{1,2,5,6}$
Xuefei Chen,$^{1,2,5,6}$\thanks{E-mail: cxf@ynao.ac.cn}
\\
$^{1}$Yunnan Observatories, Chinese Academy of Sciences (CAS), Kunming 650216, P.R. China\\
$^{2}$Key Laboratory for the Structure and Evolution of Celestial Objects, Chinese Academy of Sciences, Kunming 650011, China\\
$^{3}$Department of Physics and Astronomy, Aarhus University, Ny Munkegade 120, 8000~Aarhus~C, Denmark\\
$^{4}$Aarhus Institute of Advanced Studies (AIAS), Aarhus University, H{\o}egh-Guldbergs~Gade~6B, 8000~Aarhus~C, Denmark\\
$^{5}$University of the Chinese Academy of Sciences, Yuquan Road 19, Shijingshan Block, 100049, Beijing, China\\
$^{6}$Center for Astronomical Mega-Science, Chinese Academy of Sciences, 20A Datun Road, Chaoyang District, Bejing 100012, China
}

\date{Accepted XXX. Received YYY; in original form ZZZ}

\pubyear{2021}

\begin{document}
\label{firstpage}
\pagerange{\pageref{firstpage}--\pageref{lastpage}}
\maketitle
\begin{abstract}
Close-orbit low-mass X-ray binaries (LMXBs), radio binary millisecond pulsars (BMSPs) with extremely low-mass helium WDs (ELM He~WDs) and ultra-compact X-ray binaries (UCXBs) are all part of the same evolutionary sequence. 
It is therefore of uttermost importance to understand how these populations evolve from one specie to another. Moreover, UCXBs are important gravitational wave (GW) sources and can be detected by future space-borne GW observatories. However, the formation and evolutionary link between these three different populations of neutron star (NS) binaries are not fully understood. In particular, a peculiar fine-tuning problem has previously been demonstrated for the formation of these systems. 
In this investigation, we test a newly suggested magnetic braking prescription and model the formation and evolution of LMXBs. We compute a grid of binary evolution models and present the initial parameter space of the progenitor binaries which successfully evolve all the way to produce UCXBs. 
We find that the initial orbital period range of LMXBs, which evolve into detached NS+ELM~He~WD binaries and later UCXBs, becomes significantly wider compared to evolution with a standard magnetic braking prescription, and thus helps to relieve the fine-tuning problem. 
However, we also find that formation of wide-orbit BMSPs is prohibited for strong versions of this new magnetic braking prescription, which therefore calls for a revision of the prescription. 
Finally, we present examples of the properties of UCXBs as Galactic GW sources and discuss their detection by the LISA, TianQin and Taiji observatories. 

\end{abstract}


\section{Introduction}
\label{sec:int}
Being the end point of binary stellar evolution, binary millisecond pulsars (BMSPs) are very important for the studies 
of stellar physics and binary interactions. It is well known that BMSPs are formed from 
the evolution of low- and intermediate-mass X-ray binaries \citep[e.g.][]{bv91,tv06}. In this scenario, 
the donor star overfills its Roche lobe and transfers material and angular momentum to the neutron star (NS) which is spun up to a high spin rate. At the end of mass transfer, the system can be observed as a radio BMSP.  
Depending on the initial donor star mass and orbital period, the companion stars of BMSPs can be helium white dwarfs (He~WDs), 
carbon-oxygen WDs (CO~WDs) or low-mass dwarf stars \citep[e.g.][]{tau11,rob13,ccth13}. The typical companion stars of BMSPs are 
He~WDs, constituting some 55\% of all BMSPs (124 systems out of 227 known BSMPs with spin periods, $P<10\;{\rm ms}$, according to the ATNF Pulsar Catalogue\footnote{\url{https://www.atnf.csiro.au/research/pulsar/psrcat/} \citep{mhth05}.} in Jan.~2021).

In close-orbit BMSPs, due to efficient emission of gravitational wave (GW) radiation, the orbital periods of such detached NS+WD binaries decrease over time and the systems may reconnect and become semi-detached ultra-compact X-ray binaries \citep[UCXBs,][]{nyvt10,hie+13}, if the He~WD fills its Roche lobe within a Hubble time. The in-spiral of detached BMSPs to very tight orbits, before and after these systems become UCXBs, means that they are prime candidates for GW sources to be detected in the mHz frequency band of e.g. the LISA mission \citep{aab+17}, as well as the TianQin \citep{lcdg+16} and Taiji \citep{rgcz18} detectors, which are scheduled (or proposed) to be lunched in the next two decades. 
Hence, there is a straight evolutionary link from low-mass X-ray binaries (LMXBs), to BMSPs, UCXBs and GW sources, which will be investigated further in this paper.

Formation of BMSPs with He~WDs has been widely studied by evolving LMXBs until these systems detach from mass transfer  \citep[e.g.][]{ps88,ps89,esa98,ts99,prp02,sl12,lrp+11,itl14,jl14}.  
From these studies, it is found that there is a relationship between WD mass and orbital period \citep[e.g.][]{sav87,jrl87,rpj+95,ts99,itl14}.
It is therefore not surprising that among this type of binaries, there is a group of systems with extremely low-mass (ELM) WDs ($M_{\rm WD}\lesssim 0.20\;M_\odot$)
and short orbital periods ($P_{\rm orb}\simeq 2-9\;{\rm hr}$). These are the systems that are tight enough and thus later, due to GW radiation, evolve into UCXBs and bright GW sources. 
The details of the evolutionary sequence from LMXBs, to BMSPs, UCXBs and GW sources, however, are strongly dependent on the assumed loss of orbital angular momentum via magnetic braking (MB) and mass loss, besides GW radiation. Whereas the GW radiation in these binaries is fully deterministic \citep{pet64}, the MB is uncertain in particular. The standard prescription for MB was derived by \citet{vz81,rvj83} and is based on the empirical Skumanich law for braking the spin of low-mass (solar-type) stars via a magnetic wind \citep{sku72}.

\citet{vvp05,vvp05b} modelled the formation of UCXBs from the evolution of LMXBs through stable mass transfer, by applying the standard prescription for MB by \citet{rvj83} with a MB index of $\gamma=4$ (see Section~\ref{sec:method}).
They found that only binaries within a very narrow range of initial orbital periods and donor masses are able to produce UCXBs, 
Applying the less efficient (weaker) MB prescription by \citet{spt00}, they found it impossible to evolve any systems to the UCXB stage.

\citet{jl14} and \citet{itl14} also studied the formation of BMSPs with ELM~WD companions from the evolution of LMXBs and found that these systems can be produced from binaries with late Case~A (near the end of core hydrogen burning) or early Case~B (shortly after the onset of hydrogen shell burning) mass transfer, even if applying standard MB. 
In addition, and in analogue to the results of \citet{vvp05,vvp05b}, it was found by \citet{itl14} that only LMXBs in a very narrow range of initial orbital periods can reproduce these observed BMSPs with ELM~WDs, using standard prescriptions of orbital angular momentum losses via MB, mass loss and GWs. They therefore concluded that something needs to be modified in the standard input physics of LMXBs in order to reproduce the observed orbital period distribution of BMSPs. 

\citet{ri19} recently revisited this problem and found that the parameter space for producing BMSPs with ELM~WDs can become larger
with an alternative, and stronger, prescription of MB suggested by \citet{vih19}. This prescription will be investigated further in our work presented here.

Taken the MB induced by the coupling between the magnetic field and an irradiation-driven
wind into consideration, \citet{cp16} found that some intermediate-mass X ray binaries may also evolve into UCXBs.
On the other hand, the possible existence of significant magnetic winds in intermediate-mass stars remains an unsettled issue. 
Furthermore, the evolution of UCXBs is not well understood in detail, although significant progress has been made in the past decade 
\citep[see e.g.][who modelled analytically the evolution of UCXBs and studied the properties of their evolutionary processes]{vnv+12,vnvj12b}. However, it remained a problem (until recently, see below) that the mass transfer in compact object binaries from a WD donor to a NS accretor in UCXBs, or to a WD accretor in AM~CVn systems, had only been solved analytically, and restricted to applications of approximate zero-temperature mass-radius relations of the He~WD donor star \citep[e.g.][]{npvy01,kblk17}, or computed from hydrodynamical simulations \citep{bdc17}.

The first complete numerical computations of stable mass transfer from a He~WD to a NS were performed by \citet{stli17}, who calculated a number of coherent binary stellar evolution tracks, starting from pre-LMXBs, and evolving these to detached BMSP systems and further
on to UCXBs. The computed minimum orbital periods were as short as 5.6~min, and they were able to follow the subsequent widening of these systems until the donor stars became planets with a mass of $\sim 0.005\;M_\odot$ (5~jupiter masses). 
The properties of the computed UCXBs from their models are in good agreement with observations. 

Based on the above progress, \citet{tau18} investigated in detail the properties of these UCXB systems as GW sources. A new way to measure the NS mass was proposed, and it was suggested that some of these sources would potentially be detectable as {\em dual-line} GW sources from the combined measurements at high-frequencies by the LIGO network (from a rapidly spinning, recycled MSP with a slightly distorted shape) and at low-frequencies by LISA (from the orbital motion of the binary).    

In this first paper, we investigate various MB prescriptions \citep[based on][]{vih19} with the aim to model the full evolution from (pre-) LMXBs, to BMSPs with ELM~He~WDs and UCXBs, through stable mass transfer. We demonstrate that applying some of these new MB prescriptions enables a much wider range of initial systems to evolve all the way to the UCXB stage. At the same time, however, we show that the more strong versions of these MB prescription cannot be operating in nature, as they would exclude the formation of wide-orbit BMSPs. 
The paper is structured as follows: 
In Section~\ref{sec:method}, we introduce the numerical method and physical assumptions we used in our calculation. 
In Section~\ref{sec:res}, we present the results obtained from our simulations. 
Further discussions are given in Section~\ref{sec:diss}, including UCXBs as detecable GW sources.
We summarize our study in Section~\ref{sec:con}.

\section{Method and assumptions}
\label{sec:method} 
In this work, we use the stellar evolution code \emph{Modules for Experiments in Stellar Astrophysics} 
\citep[\texttt{MESA} version 12115,][]{pbd+11,pcab+13,pmsb+15,psbb+18,pss+19} 
to model the binary evolution. We treat the NS accretor as a point mass with an assumed initial mass of 
either $M_{\rm NS}=1.30\;M_{\odot}$ or $M_{\rm NS}=1.80\;M_{\odot}$. The donor stars are initially zero-age main-sequence (ZAMS) stars and their masses, $M_2$
range from $1.0\;M_{\odot}$ to $1.75\;M_{\odot}$. Their initial orbital periods range from around 0.40\;d to 1000\;d, and their chemical abundance is assumed to have a metallicity of $Z=0.02$ (with a hydrogen abundance, $X = 0.70$, and a helium abundance, $Y = 0.28$). 
In our models, we further assume a mixing-length parameter of $\alpha = l/H_{\rm p} = 2.0$ and overshooting is not considered. Regarding the stellar wind, 
we adopt the wind prescription from \citet{rei75}:
\begin{equation}
    \dot{M}_{\rm 2,wind} = -4\times10^{-13}\;M_\odot\,{\rm yr}^{-1}\quad \eta\,\left(\frac{R_2}{R_\odot}\right) \left(\frac{L_2}{L_\odot}\right) \left(\frac{M_\odot}{M_2}\right)\;,
\end{equation}
where we chose a scaling factor of $\eta =1$. Here, $R_2$ and $L_2$ refer to the radius and luminosity of the donor star.

In our calculations, we also take the rotation of the donor stars into consideration. 
The initial angular rotation velocities of the donor stars are assumed to be synchronized with the initial orbital angular velocities of the binaries, $\omega_{\rm orb}$. 
We consider various rotationally-induced instabilities that result during the evolution due to chemical mixing and transport of angular momentum \citep[see][for more details]{hlw00}. 
With respect to the chemical mixing, we include the mixing induced by secular shear instability, Eddington-sweet circulation, dynamical shear instability and Goldreich-Schubert-Fricke instability. We assume a mixing efficiency factor, $f_{\rm c} = 1/30$ \citep{hlw00}. 
We include the transport of angular momentum induced by the Spruit-Tayler dynamo \citep{spr02,hws05}. 
The inhibiting effect of chemical gradients on the mixing process is regulated by the parameter $f_{\rm \mu}$ and 
we set it to be $0.05$ following \citet{hlw00}. 

Regarding the angular momentum loss, we include orbital angular momentum loss due to mass loss, GWs, magnetic braking and other spin-orbit couplings \citep[e.g.][]{tv06}.
The orbital angular momentum loss due to GW radiation can be computed with the following formula \citep{ll71}:
\begin{equation}
	\frac{{\rm d}J_{\rm gw}}{{\rm d}t} = -\frac{32}{5}\frac{G^{7/2}}{c^{5}}\frac{M^2_{\rm NS}M^2_2(M_{\rm NS}+M_2)^{1/2}}{a^{7/2}}\;,
\end{equation}
where $G$ is the gravitational constant, $c$ is the speed of light in vacuum, $a$ is the binary separation, $M_{\rm NS}$ and $M_{2}$ are the mass of the NS and the companion star, respectively.

To compute the loss of orbital angular momentum due to MB, we adopt the newly suggested prescription from \citet{vih19}:
\begin{equation}
    \frac{{\rm d}J_{\rm mb}}{{\rm d}t} = \frac{{\rm d}J_{\rm mb,Sk}}{{\rm d}t}\;\left(\frac{\omega_2}{\omega_\odot}\right)^{\beta}\left(\frac{\tau_{\rm conv}}{\tau_{\rm \odot,conv}}\right)^{\xi}\left(\frac{\dot{M}_{\rm 2,wind}}{\dot{M}_{\rm \odot,wind}}\right)^{\alpha}\;,
\label{eq:mb_vih}
\end{equation}
where
\begin{equation}
	\frac{{\rm d}J_{\rm mb,Sk}}{{\rm d}t} = -3.8 \times 10^{-30}\;M_2\,R_\odot^4\,\left(\frac{R_2}{R_\odot}\right)^{\gamma_{\rm mb}}\omega_2^{3}\quad{\rm dyne\,cm}\;,
\end{equation}
is the magnetic braking law derived by \citet{rvj83} 
and where $\gamma_{\rm mb} = 4$ is for the standard Skumanich MB law \citep{vz81}, $\omega_2$ is the angular velocity of the donor star and $\omega_{\odot} = 3.0 \times 10^{-6}\;{\rm s}^{-1}$, $\tau_{\rm conv}$ is the turnover time of convective eddies and $\tau_{\rm \odot,conv} = 2.8 \times 10^{6}\;{\rm s}$, $\dot{M}_{\rm 2,wind}$ is the wind mass-loss rate of the donor star and $\dot{M}_{\rm \odot,wind} = 2.54 \times 10^{-14}\;M_{\odot}\,{\rm yr}^{-1}$ \citep{co06}.

\citet{vih19} considered four cases (here denoted MB1 to MB4) of different origin and strength in their prescription of MB:
\begin{equation}
  (\beta,\,\xi,\,\alpha) = \, \left\{ 
  \begin{array}{ll}
  (0,\,0,\,0) & \mbox{\hspace{0.1cm} MB1 --- standard MB}\\
  (0,\,2,\,0) & \mbox{\hspace{0.1cm} MB2 --- convection-boosted MB}\\
  (0,\,2,\,1) & \mbox{\hspace{0.1cm} MB3 --- intermediate MB}\\
  (2,\,4,\,1) & \mbox{\hspace{0.1cm} MB4 --- wind-boosted MB}\\
  \end{array}
  \right.
  \label{eq:MB-laws}
\end{equation}
They discussed different choices of values of $\beta$, $\xi$ and $\alpha$, and compared the results with observations of LMXBs. 
They found that models with $\beta = 0$, $\xi = 2$ and $\alpha = 1$ (MB3) can better reproduce the observed orbital characteristics of these X-ray binaries (see their fig.~7) compared to the other prescriptions. 
Therefore, here in our work, we adopt the MB3 prescription as our default test model, but we will also discuss in Section~\ref{subsec:dis_mb} the influence of different choices of the parameters $(\beta,\,\xi,\,\alpha)$ on our results. 

If $\beta = \xi = \alpha = 0$, the MB in equation~(\ref{eq:mb_vih}) becomes the standard Skumanich law (MB1) derived by \citet{vz81} and \citet{rvj83}. 
The application of MB3, compared to the standard prescription of MB1, includes the additional scaling factors that account for the effects of wind mass-loss and turnover time for convection, leading to a stronger MB. This stronger MB prescription will significantly reduce the binary separation on a shorter timescale compared to standard MB \citep{vih19}. 
It is worth noting that the factor $\tau_{\rm conv}$ can be smaller than $\tau_{\rm \odot,conv}$, depending on the evolutionary phase of the star. Therefore, if $\beta = \alpha = 0$ are adopted, this sole MB effect of $\xi \ne 0$ can lead to weaker MB than MB1. 
In addition, we assume that MB only operates for binaries with donor masses, $M_2 \le 1.50\;M_\odot$. This is because these low-mass stars have convective envelopes which are thought to drive a dynamo and produce a significant B-field, while more massive stars have no significant convective envelopes \citep[however, see][for discussions on the anomalously high B-fields associated with Ap and Bp stars]{jrp06,che17}.

In our calculation, the scheme proposed by \citet{rit88} is adopted to compute the mass-transfer rate and an isotropic re-emission model of mass transfer is adopted \citep[see e.g.][for a review]{tv06}. In this model, the material flows over from the secondary, in a conservative way, to the NS via Roche-lobe overflow (RLO), and from here a fraction of this material is re-emitted with the specific orbital angular momentum of the NS. 
We assume that the accretion efficiency of the NS is 0.30, i.e. 30\% of the material transferred from the donor star is accreted by the NS \citep{ts99,akk+12}, 
and we neglect direct wind mass loss from the donor star and formation of circumbinary coplanar toroid in the orbital angular momentum budget. 
In addition, the accretion rate of the NS is limited by Eddington mass-accretion rate, which is given by: 
\begin{equation}
	\dot{M}_{\rm Edd} = \frac{4\pi GM_{\rm NS}}{\eta \,0.20(1+X) c}.
\end{equation}
Here $X$ is the hydrogen mass fraction of the accreted material. The ratio of gravitational mass to baryonic mass of accreted material is 
assumed to be 0.85, i.e. $\eta = 0.15$ \citep[e.g.][]{lp07}. 

Regarding the angular momentum loss due to spin-orbit coupling, we adopt the prescription from \citet{pmsb+15}. The time evolution of the 
angular frequency of the donor star is:  
\begin{equation}
\frac{{\rm d}\omega_{\rm j}}{{\rm d}t} = \frac{\omega_{\rm orb}-\omega_{\rm j}}{\tau_{\rm sync}},
\end{equation}
where $\omega_{\rm j}$ is the angular frequency at the face of cell $j$, $\tau_{\rm sync}$ is the synchronization time and computed with 
the formula of tidal effects from \citet{htp02}.

\section{Results}
\label{sec:res}

\subsection{Examples of binary evolution}
\label{subsec:bin_ex}

\begin{figure}
    \includegraphics[width=0.96\columnwidth]{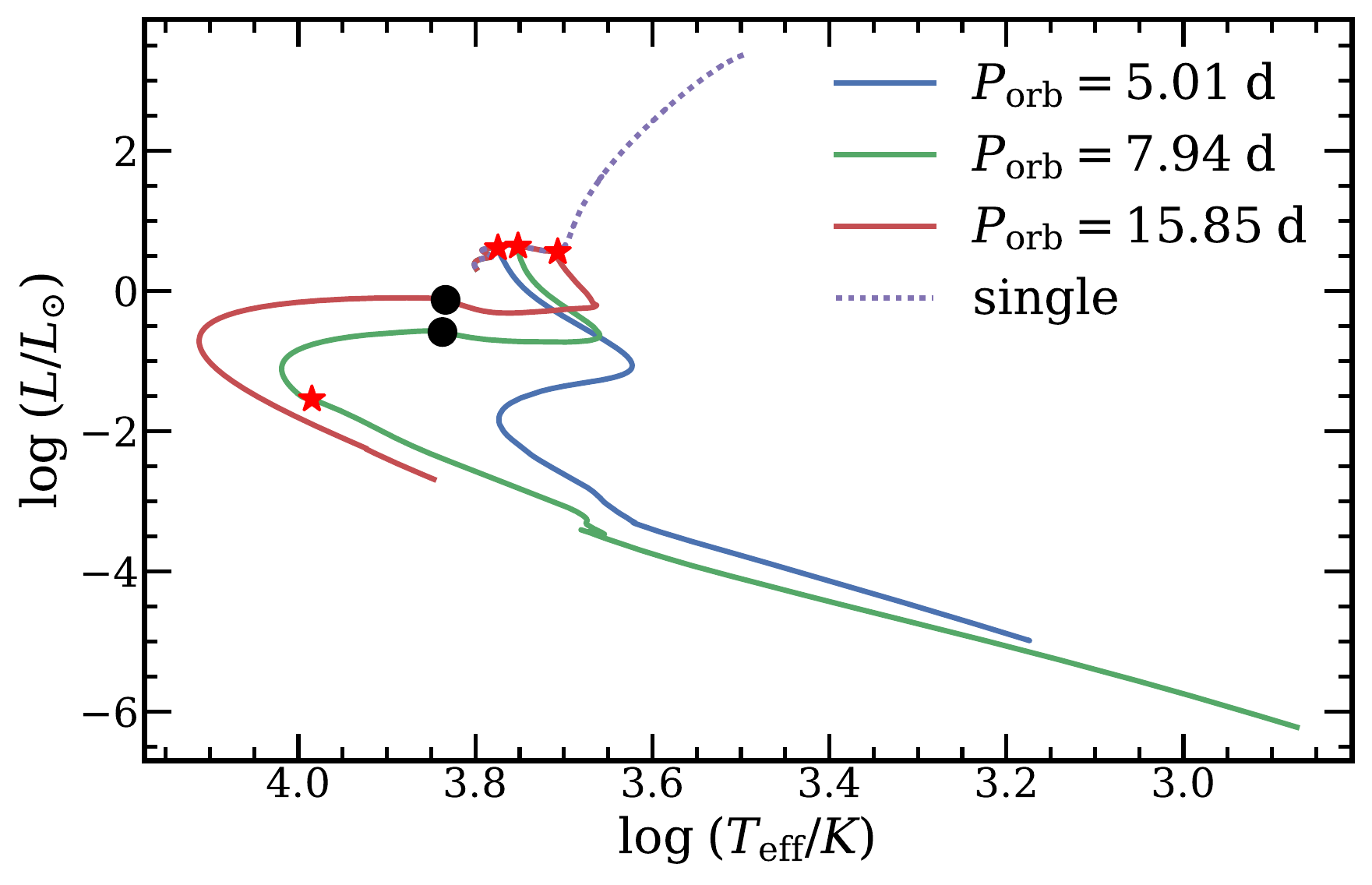}
    \includegraphics[width=0.96\columnwidth]{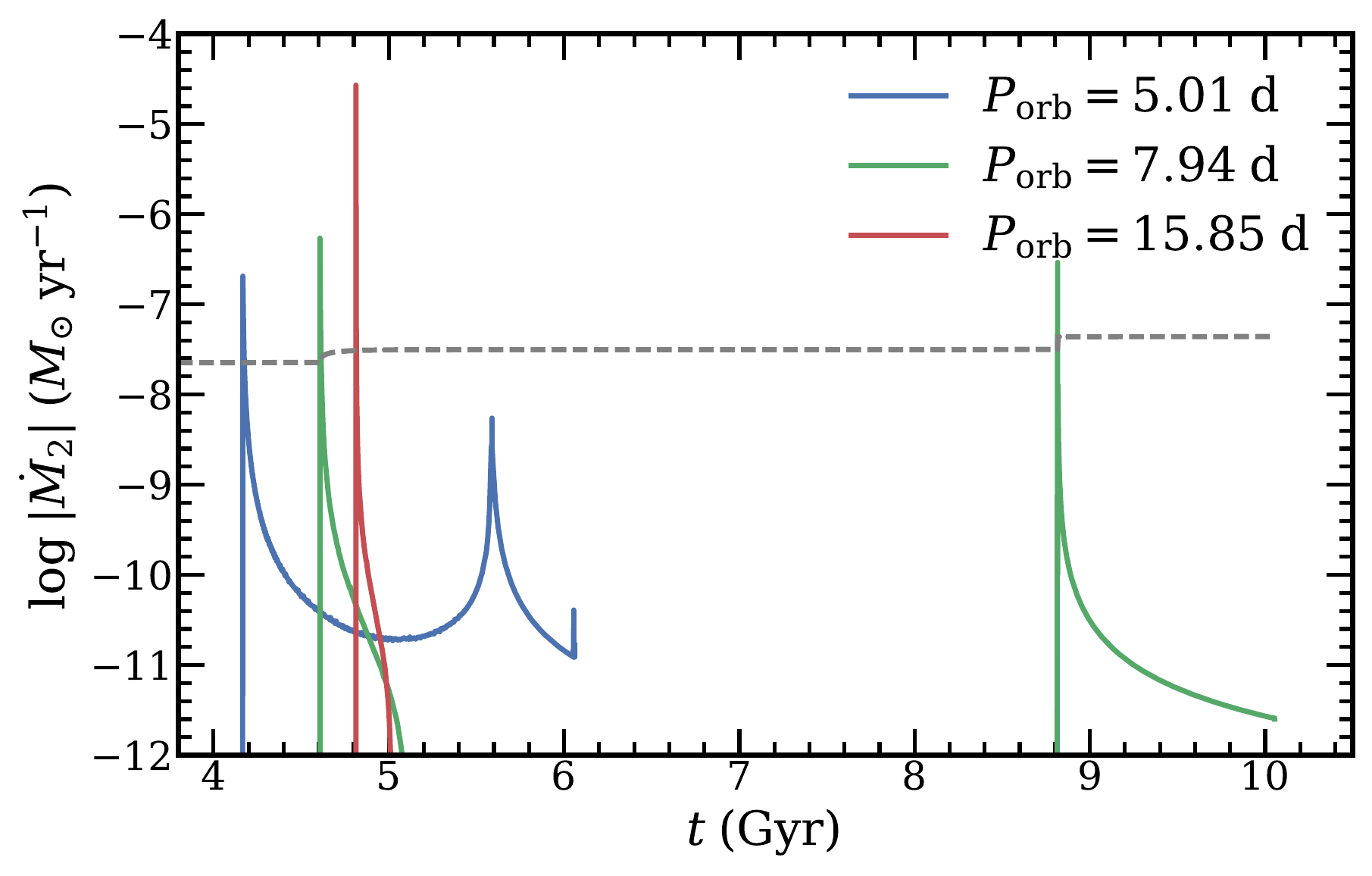}
    \includegraphics[width=0.96\columnwidth]{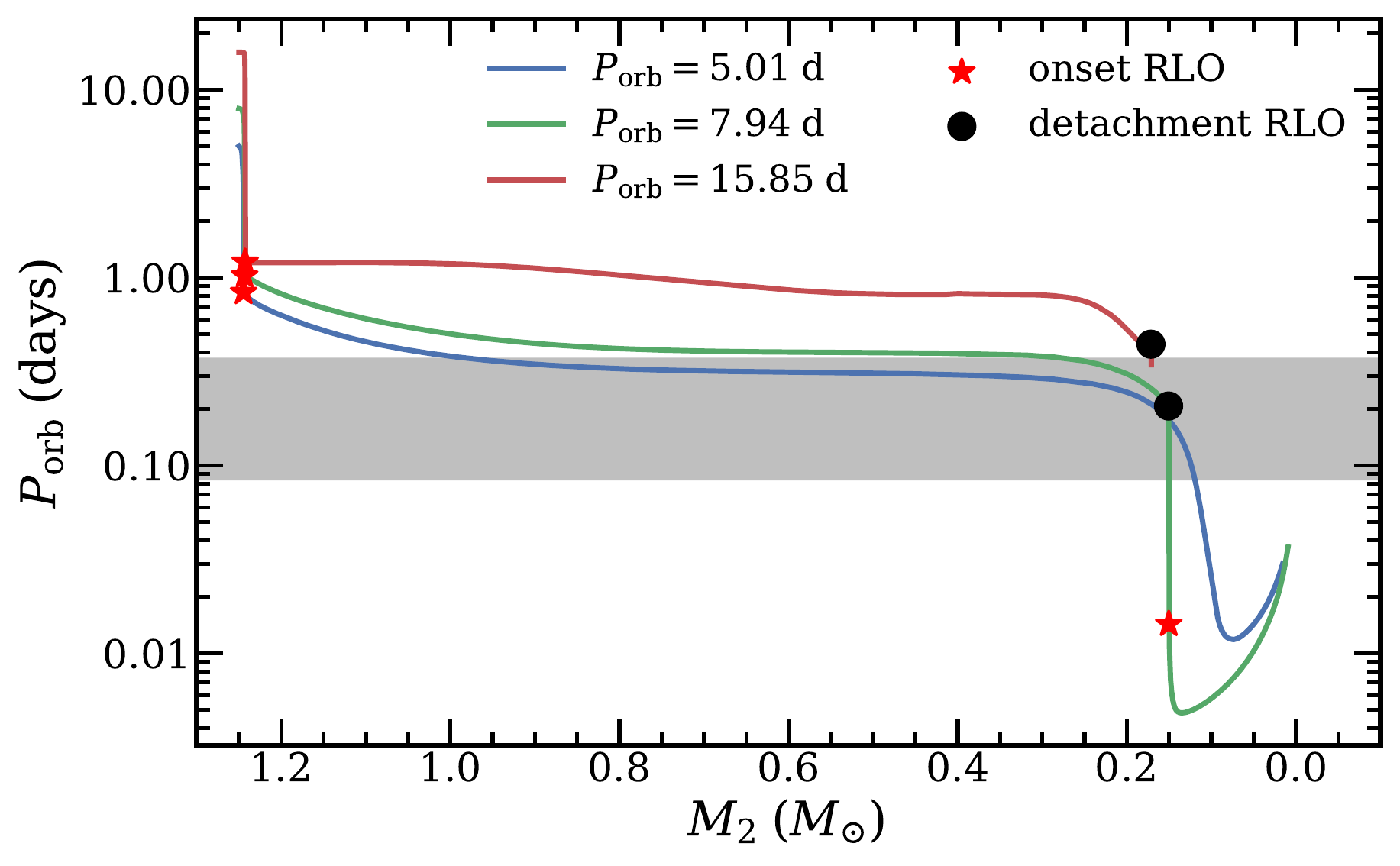}
    \caption{Three examples of binary evolution with initial orbital periods of $5.01\;$d (blue lines), $7.94$\;d (green lines) and $15.85$\;d (red lines) and MB3 prescription. The initial NS and donor masses in all cases are $1.30\;M_{\odot}$ and $1.25\;M_{\odot}$, respectively. 
    The upper panel is the HR-diagram. The evolutionary track of a single star with $1.25\;M_{\odot}$ is also plotted (dotted line) for comparison. The middle panel shows the evolution of mass-transfer rate as a function of time. The grey dashed line shows the Eddington accretion rate for the model with an orbital period of $P_{\rm orb} = 7.94\;$d. The lower panel shows the evolution of orbital period as a function of decreasing donor mass. The red stars denotes the beginning of RLO, and black circles represent the end of RLO. The grey shaded region shows the orbital period range ($2-9\;{\rm hr}$) of the observed BMSPs with ELM WDs \citep[see][]{itl14}.
    }
    \label{fig:bin_ev_ex}
\end{figure}

\begin{figure}
    \includegraphics[width=1.10\columnwidth]{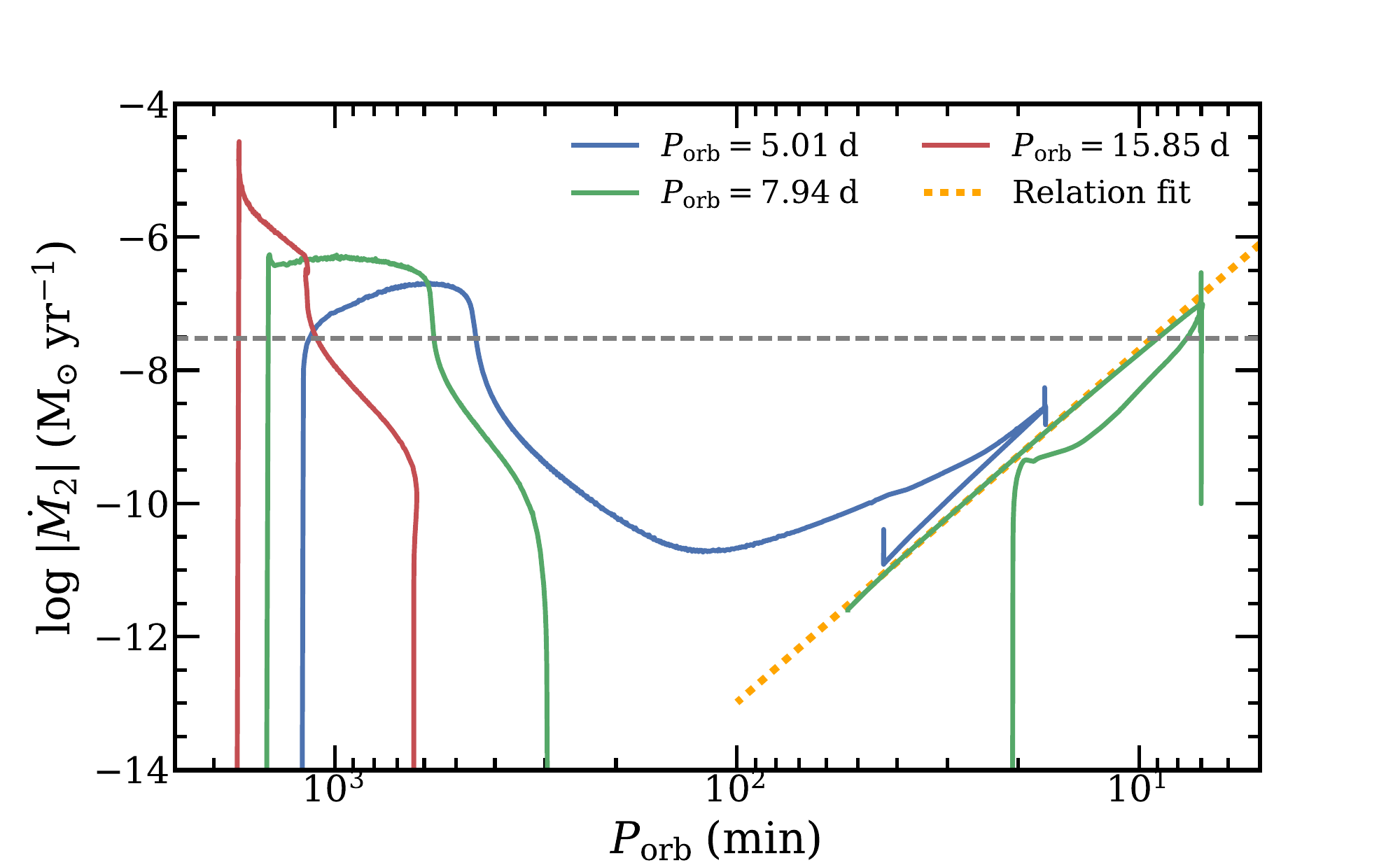}
    \caption{Evolution of mass-transfer rate as a function of orbital period for the three binaries plotted in Fig.~\ref{fig:bin_ev_ex}. 
    The evolution is from left to right, until reaching the orbital period minimum (blue and green tracks) at which point the orbits expand again (i.e. evolve to the left). Near the orbital period minimum at $\sim 7\;{\rm min}$ (green track) and $\sim 17\;{\rm min}$ (blue track), as well as around 45~min for the latter, vertical lines occur due to numerical noise in our computations. 
    The grey dashed line represents the Eddington accretion rate $\dot{M}_{\rm Edd} \simeq 3.0 \times 10^{-8}\;M_\odot \,{\rm yr}^{-1}$. 
    The orange dotted line represents the fitted relation of the common declining branch of UCXBs from \citet{stli17}.
    }

    \label{fig:porb_lgmdot_ex}
\end{figure}

\begin{figure}
    \centering
    \includegraphics[width=\columnwidth]{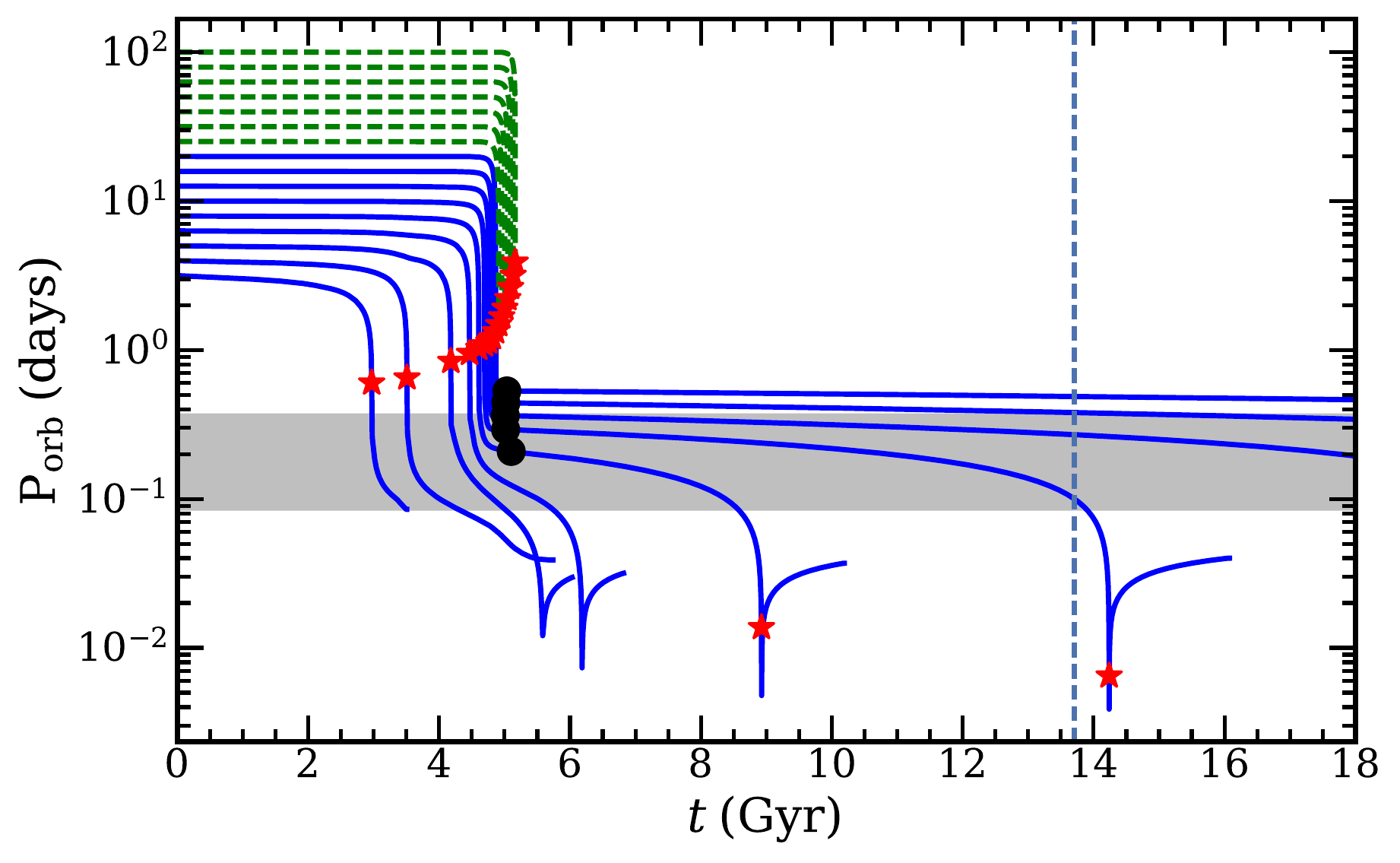}
    \caption{Evolution of orbital period as a function of time for systems with $M_{\rm NS} = 1.30\;M_\odot$, $M_2 = 1.25\;M_\odot$ and different initial orbital periods. All tracks are calculated with the MB3 prescription. 
    The red stars and black circles represent the beginning and end of mass transfer, respectively. The vertical dashed line represents the Hubble time (i.e. 13.7~Gyr). The green dashed tracks in the wider orbits represent binary systems which undergo dynamically unstable mass transfer (see text). 
    The grey-shaded region shows the orbital period range ($2-9\;{\rm hr}$) of the observed NS+ELM~WDs which will become UCXBs. See Fig.~\ref{fig:ele_bin_ex} for a zoom-in of such solutions.
    }
    \label{fig:t_porb_diff_porb}
\end{figure}

\begin{figure}
    \includegraphics[width=\columnwidth]{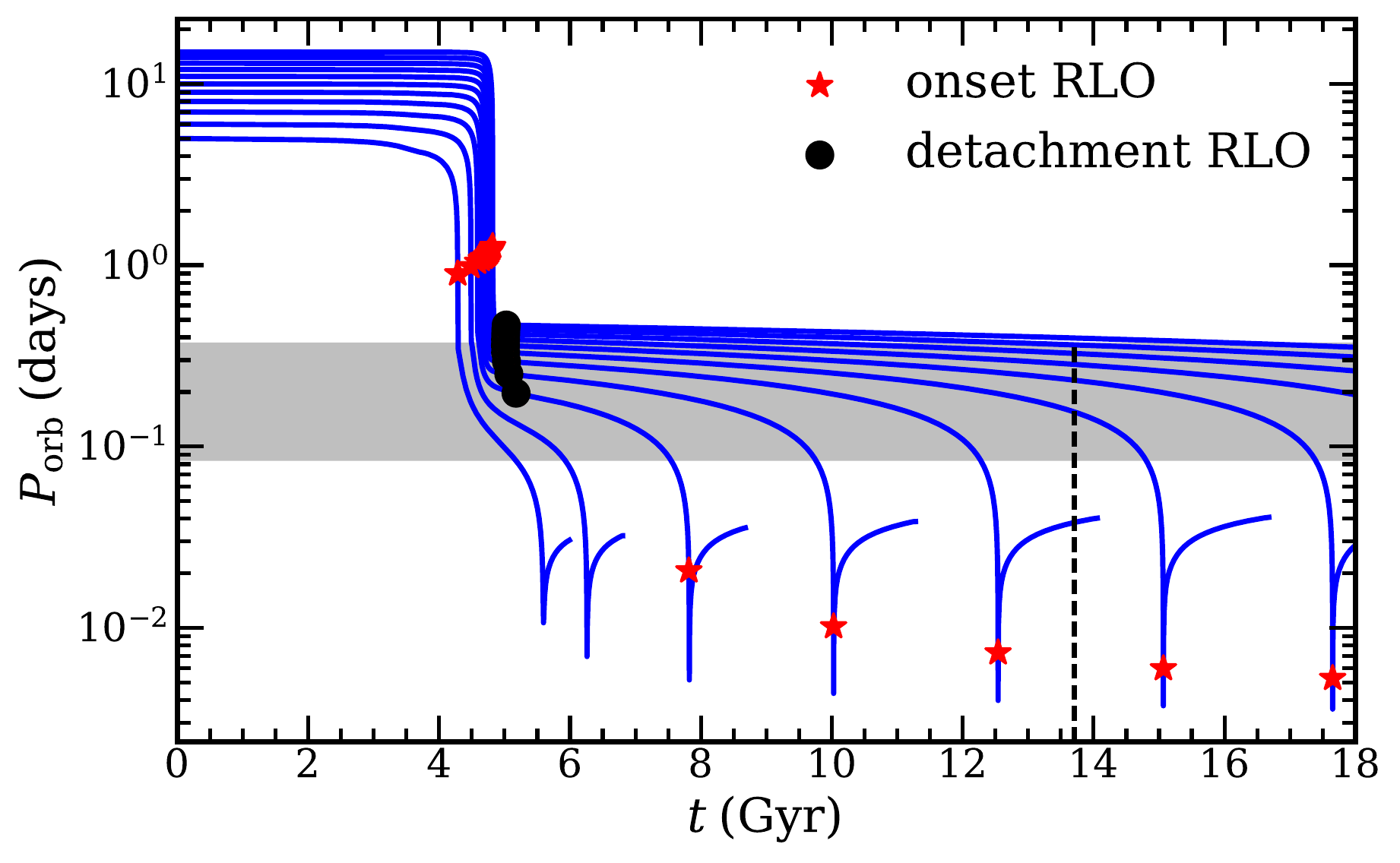}
    \caption{Similar to Fig.~\ref{fig:t_porb_diff_porb}, but with a higher resolution for a selected initial orbital period range between 5.0 and 15.0~d with a step of 1.0~d. 
    }
\label{fig:ele_bin_ex}
\end{figure}

In Fig.~\ref{fig:bin_ev_ex}, we present the evolution of three binaries with an initial NS mass, $M_{\rm NS} = 1.30\;M_{\odot}$, donor star mass, $M_2 = 1.25\;M_{\odot}$, and orbital periods of $P_{\rm orb} = 5.01$, 7.94, and 15.85\;d. 
All binaries are calculated with the prescription MB3. 
The three panels display the HR-diagram (upper), the mass-transfer rate, $|\dot{M}_2|$ as a function of stellar age (middle), and orbital period, $P_{\rm orb}$ as a function of decreasing donor star mass (lower).
Note, the computations of the evolutionary tracks for the system with $P_{\rm orb} = 5.01\;{\rm d}$ and 7.94~d do not converge due to numerical issues 
once the donor star becomes a He-rich dwarf star with a mass, $M_2 \la 0.015\;M_\odot$ and $M_2 \la 0.009\;M_\odot$, respectively.
All of these three systems start mass transfer when their donor stars are in the Hertzsprung gap (HG), i.e. Case~B RLO, but their evolution is quite different.
In all three examples, the orbital periods decrease significantly, down to about 1~d (due to the relatively efficient MB prescription MB3) before RLO is initiated.
We now discuss the differences between the evolution of these three systems.

The system with smallest initial orbital period never becomes detached from RLO and therefore does not produce a detached He~WD.
Initially, when the secondary (donor) star initiates mass transfer, it has a small He~core with mass $\sim 0.07\;M_\odot$. 
The core does not increase significantly during the mass transfer. 
At $t=5.58\;{\rm Gyr}$, the secondary star finally loses its H~envelope completely and the naked (still barely degenerate) He~core is exposed while the RLO of this UCXB source continues with increasing $|\dot{M}_2|$. 
Shortly thereafter, around the minimum orbital period ($\sim 17\;{\rm min}$), the secondary star becomes fully degenerate. The orbit then starts to widen (increasing $P_{\rm orb}$) as the mass transfer continues, although now at a reduced rate.

The second system with an initial period, $P_{\rm orb} = 7.94\;{\rm d}$ has a $0.108\;M_\odot$ He~core at 
the beginning of the first mass-transfer phase. During the RLO, the He~core grows and its mass increases to
$\sim 0.12\;M_\odot$ (the total mass is $\sim 0.15\;M_\odot$ and the H-rich envelope mass is around $0.03\;M_\odot$) by the time it detaches as a He~WD with $P_{\rm orb}=5.0\;{\rm hr}$. Thereafter, due to GW radiation, the binary separation 
decreases over time and the He~WD fills its Roche lobe after a detached phase of 3.74~Gyr. During this epoch, the system is expected to be visible as a radio BMSP, depending on beaming direction. The system then evolves into an UCXB once the He~WD initiates a second phase of RLO at $P_{\rm orb}\simeq 20\;{\rm min}$. The minimum orbital period is 6.9~min.

In the third system, the binary separation is relatively large after the donor loses its H~envelope and terminates RLO. This happens when $M_2=0.17\;M_\odot$ and $P_{\rm orb}=10.2\;{\rm hr}$. Thereafter, the angular momentum loss due to GW radiation is not strong enough to make the NS+WD system semi-detached (and start another phase of RLO) within the age of the Universe.

In Fig.~\ref{fig:porb_lgmdot_ex}, we show the evolution of mass-transfer rate as a function of orbital period for the above three binaries. The minimum orbital period for the system with an initial orbital period, $P_{\rm orb} = 7.94\;{\rm d}$ is smaller than that of the system with $P_{\rm orb} = 5.01\;{\rm d}$. 
This is because the $P_{\rm orb} = 7.94\;{\rm d}$ system (green), after the LMXB phase, is detached for a time interval of 3.74~Gyr, which causes the He~WD to fully contract and settle on the cooling track before refilling its Roche lobe, and thus having a radius of only $R_2=0.028\;R_\odot$ at the orbital period minimum. In comparison, the $P_{\rm orb} = 5.01\;{\rm d}$ system (blue) never detaches from the LMXB phase, whereby the donor star remains somewhat bloated in radius while gradually losing mass and reaching the minimum orbital period with $R_2=0.043\;R_\odot$. This explanation was also found by \citet{stli17} who also argued that when the donor stars in such UXCBs become degenerate, their evolutionary tracks start to converge along the same path (orange dotted line). For further discussions on cooling and contracting of ELM~WDs, see e.g. \citep{amc13,itla14,imt+16}.

In Fig.~\ref{fig:t_porb_diff_porb}, we present the evolution of orbital periods as a function of time for binaries with an initial NS mass, $M_{\rm NS} = 1.30\;M_\odot$, a donor star mass, $M_2 = 1.25\;M_\odot$, and different initial orbital periods and evolving according to the MB3 prescription. In our calculations, we find binary systems will have dynamically unstable mass transfer if their initial orbital periods are larger than about 25~d, which are shown with green dashed lines in the Fig.~\ref{fig:t_porb_diff_porb}. 
For such binaries, our calculations stop because of numerical convergence problems shortly after the onset of RLO due to very large values of $|\dot{M}_2|$.
The mass-transfer rates reach values of $|\dot{M}_2| > 10^{-4} \;M_\odot \,{\rm yr}^{-1}$ which would most likely lead to a common-envelope (CE) phase \citep{ijc+12}. For binaries with different donor masses, we find similar results. 
Thus we conclude that no wide-orbit binary MSPs can be produced when applying the MB3 prescription\footnote{If we take donor stars with $M_2 > 1.50\;M_\odot$ into account, we were able to produce BMSPs with final orbital periods up to 59 days -- although still far below the observational upper limit.}. We notice that this is not the case for the calculations presented by \citet{ri19}. The reason for this discrepancy is unknown and should be resolved. 

In order to better constrain the possible range of initial orbital periods within which binaries can evolve into detached\footnote{In our calculations, we assumed that a binary system is a detached system if its mass-transfer rate is smaller than $1.0\times10^{-14}\;M_\odot\,{\rm yr}^{-1}$.} NS+He~WDs binaries with orbital periods, $P_{\rm orb} \sim 2-9\;{\rm hr}$ (i.e. BMSPs which will evolve into UCXBs within a Hubble time),  
we computed the evolution of a grid of X-ray binaries with a smaller step in initial orbital periods.
This is shown in Fig.~\ref{fig:ele_bin_ex}, which displays the evolution of similar systems with initial orbital periods between $5.0-15.0\;{\rm d}$ with a step size of $1.0\;{\rm d}$.

It is clearly seen from Fig.~\ref{fig:ele_bin_ex} that systems with larger initial orbital periods (prior to the LMXB phase) end up with systematically smaller orbital periods at the onset of the UCXB phase (when the He~WD initiates RLO, see red stars). This result was also found by \citet{stli17} and is due to the He~WD masses and binary separations being larger at the end of first mass-transfer phase (i.e. the LMXB phase). Hence, it takes longer time for these detached NS+He~WD systems to become semi-detached, and therefore the He~WDs have longer time to cool down and reduce their radii before the onset of RLO in the UCXB phase. The minimum orbital periods of these systems during the UCXB phases are roughly between $5-9\;{\rm min}$. This plot can be directly compared with fig.~2 in \citet{itl14} and fig.~1 in \citet{stli17}.

From this comparison, we find that the initial orbital period range of binary systems which can produce BMSPs with ELM~WDs in the desired range of 2 to 9~hr is considerably larger (between $7.0\;{\rm d}$ and $14.0\;{\rm d}$) if we adopt the MB3 prescription from \citet{vih19}, compared to the default MB1 prescription applied e.g. in \citet{itl14} with which we find that the initial orbital period range is very narrow (between $2.985-3.005\;{\rm d}$).
Therefore, applying the MB3 prescription instead of the MB1 prescription, the previously mentioned fine tuning-problem is relieved and many more LMXBs can make it to the UCXB stage. 

We conclude that although the MB3 prescription relieves the fine-tuning problem in producing UCXBs, it introduces a new problem: namely, that the many observed radio BMSPs in wide orbits cannot be produced using this prescription.

\subsection{Initial binary parameter space}
\label{subsec:para_spa}
\begin{figure}
\includegraphics[width=\columnwidth]{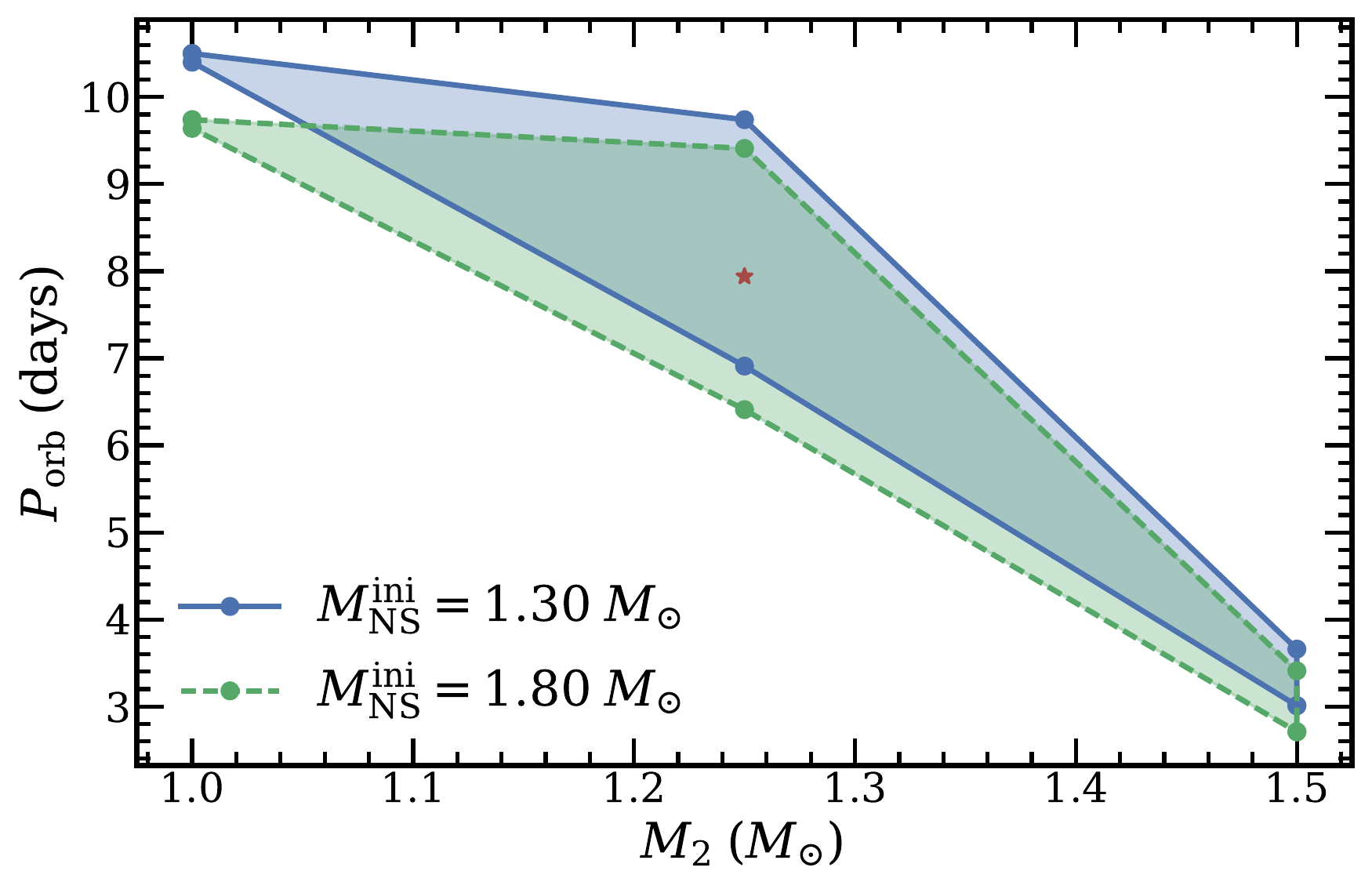}
\caption{Initial parameter space of LMXBs which can evolve into detached NS+He~WDs and then UCXBs, when applying the MB3 prescription. 
	The initial NS masses for the blue and green regions are $1.30\;M_\odot$ and $1.80\;M_\odot$, respectively. 
	The point with the red star symbol represents the system with $P_{\rm orb} = 7.94\;{\rm d}$ in Fig.~\ref{fig:bin_ev_ex}.
	}
\label{fig:ini_par}
\end{figure}

In order to investigate what kind of X-ray binaries will evolve into detached NS+He~WDs and then UCXBs, we compute a grid 
of X-ray binary evolution. The initial NS mass is assumed to be $1.30\;M_\odot$ or $1.80\;M_\odot$. The donor 
star mass ranges from 1.0 to $1.75\;M_\odot$ with a step of $\Delta M = 0.25\;M_\odot$. The initial orbital period 
ranges from $\log (P_{\rm orb}/{\rm days}) = -0.10$ to $\log (P_{\rm orb}/{\rm days}) = 3.0$ with a step 
of $\Delta \log (P_{\rm orb}/{\rm days}) = 0.10$. From this grid, we find the possible initial orbital period range. 
To obtain a higher resolution, we subsequently decrease the step of initial orbital period to $\Delta P_{\rm orb} = 0.10\;{\rm d}$ (or even $0.02\;{\rm d}$ in some cases) to get more precise boundaries. We stop the calculation
when the evolutionary time is larger than the age of the Universe (i.e. the Hubble time of 13.7~Gyr), or if the He~WD mass becomes too small, leading to numerical convergence problems.

In Fig.~\ref{fig:ini_par}, we present with shaded regions the initial parameter space of LMXBs which can evolve into detached 
NS+He~WDs and then UCXBs within a Hubble time. We notice that the  parameter space for different NS masses is quite similar.
The binaries with initial orbital periods {\em below} the shaded regions start mass transfer on the main sequence (Case~A RLO) and have no or a very small He~core at the beginning of mass transfer. 
These systems do not evolve into detached NS+He~WDs. On the other hand, binaries with initial orbital periods
{\em above} the shaded parameter space may evolve into detached NS+He~WDs, but do not produce UCXBs. 
Examples of such evolution is shown in Fig.~\ref{fig:ele_bin_ex}, where we see that there are some models which evolve into detached NS+WD systems, but not UCXBs within a Hubble time. 
For even larger initial orbital periods, the RLO becomes dynamically unstable, as displayed with green dashed tracks in  Fig.~\ref{fig:t_porb_diff_porb}.

We note that, we do not find any systems with donor masses, $M_2 > 1.50\;M_\odot$ which evolve into detached NS+He~WDs and then UCXBs. This result is different from that of \citet{clw20} who find that some intermediate-mass X-ray binaries can evolve into UCXBs. 
This is mainly because the latter authors allow for MB to operate for stars with $M_2 > 1.50\;M_\odot$. Therefore, the orbital angular momentum of these system can be effectively reduced. 
In our simulations, the systems with  $M_2 > 1.50\;M_\odot$ can only evolve into detached NS+He~WD systems with He~WD masses $M_{\rm WD} > 0.20\;M_\odot$ and in relatively wide orbits that will not become semi-detached within a Hubble time.

\section{Discussion}
\label{sec:diss}

\subsection{Influence of magnetic braking}
\label{subsec:dis_mb}
Modelling MB in binaries is still a challenging topic. Besides the tidal interaction, even the torque acting on an isolated star with a magnetic stellar wind is currently an issue of debate \citep{gdg+20}.
\citet{vih19} have proposed a number of different braking laws (equation~\ref{eq:MB-laws}) based on a set of free parameters.
We will now discuss the influence of these parameters on our results. 

\begin{figure}
\centering
\includegraphics[width=\columnwidth]{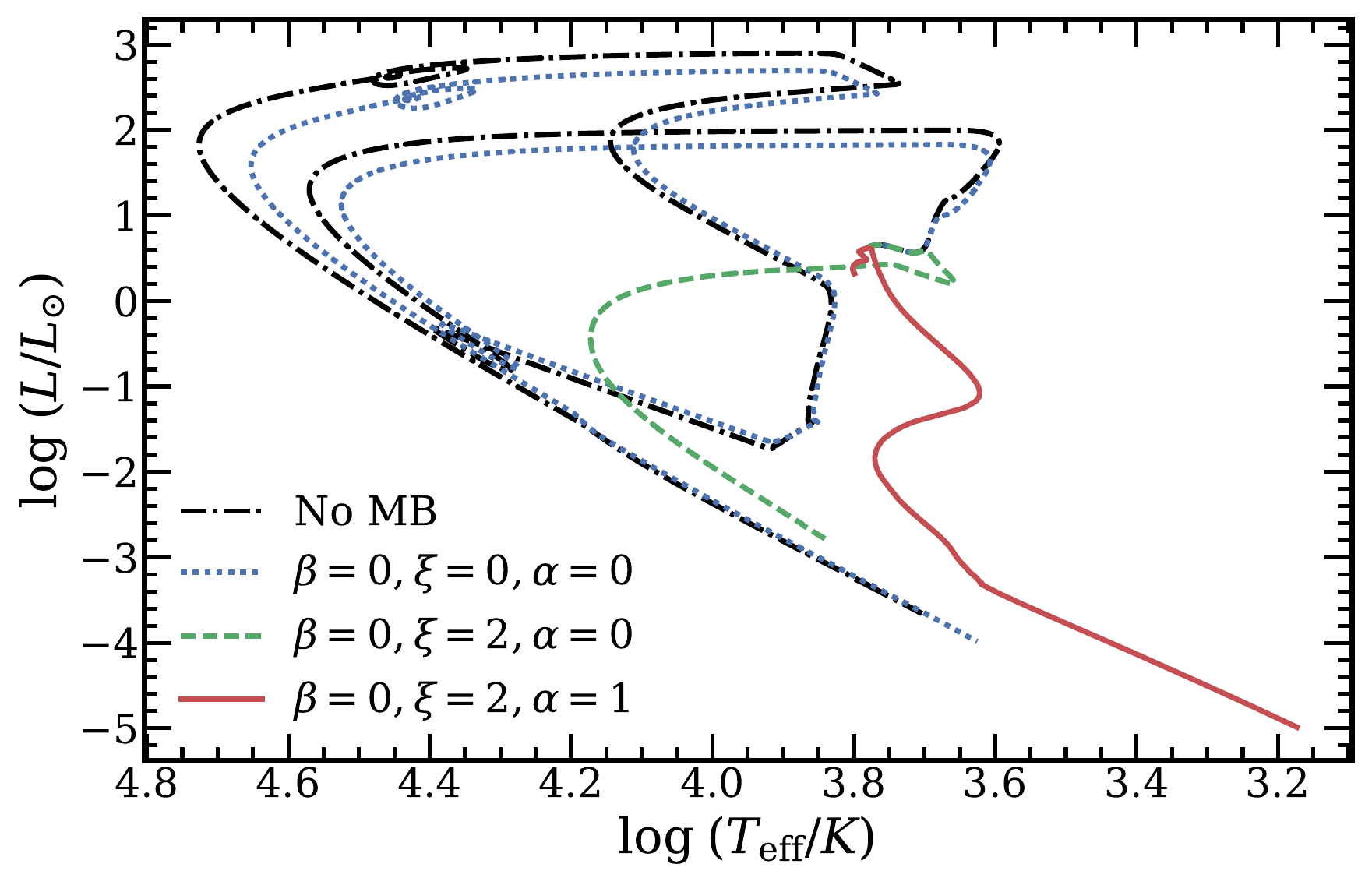}
\includegraphics[width=\columnwidth]{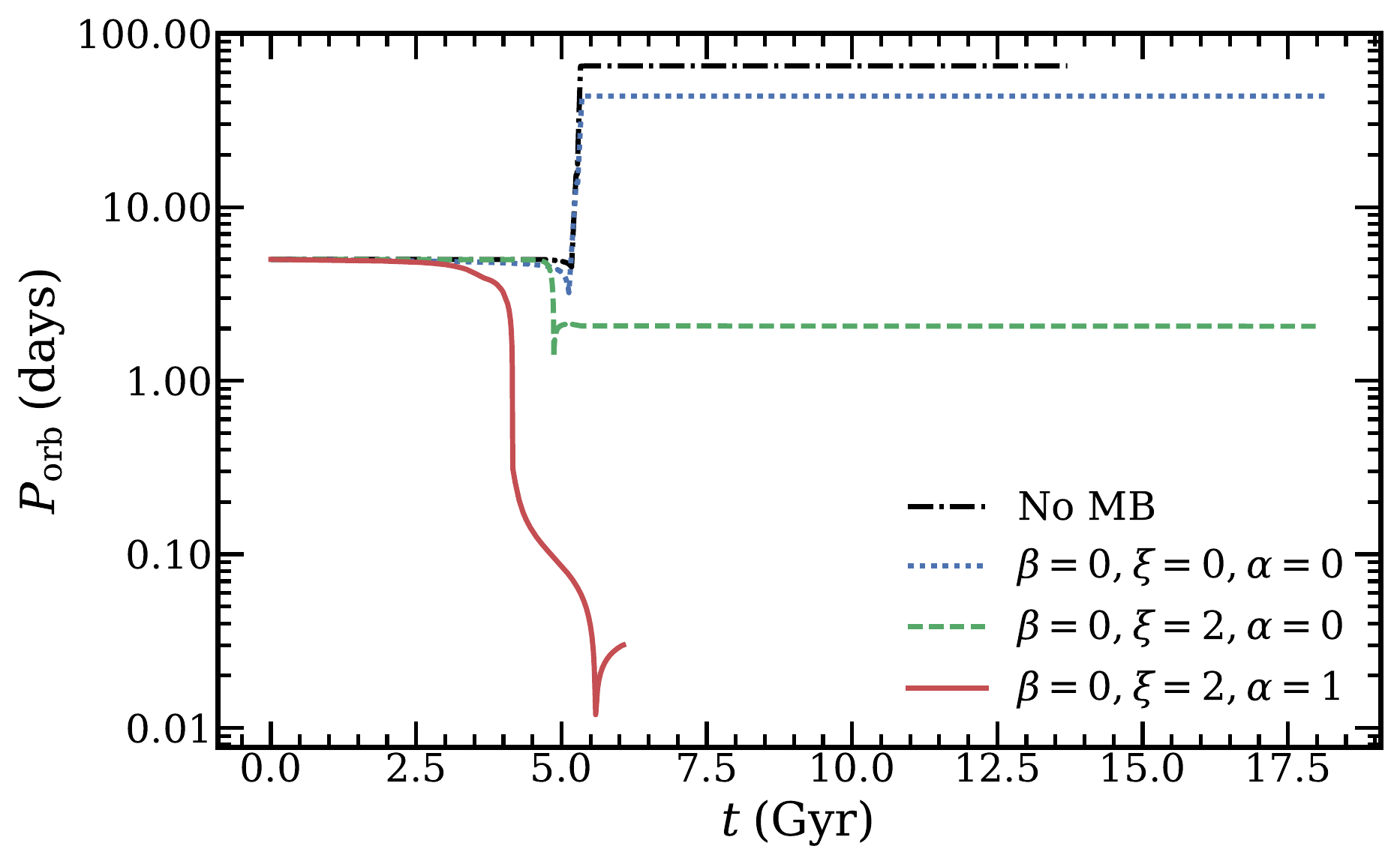}
\includegraphics[width=\columnwidth]{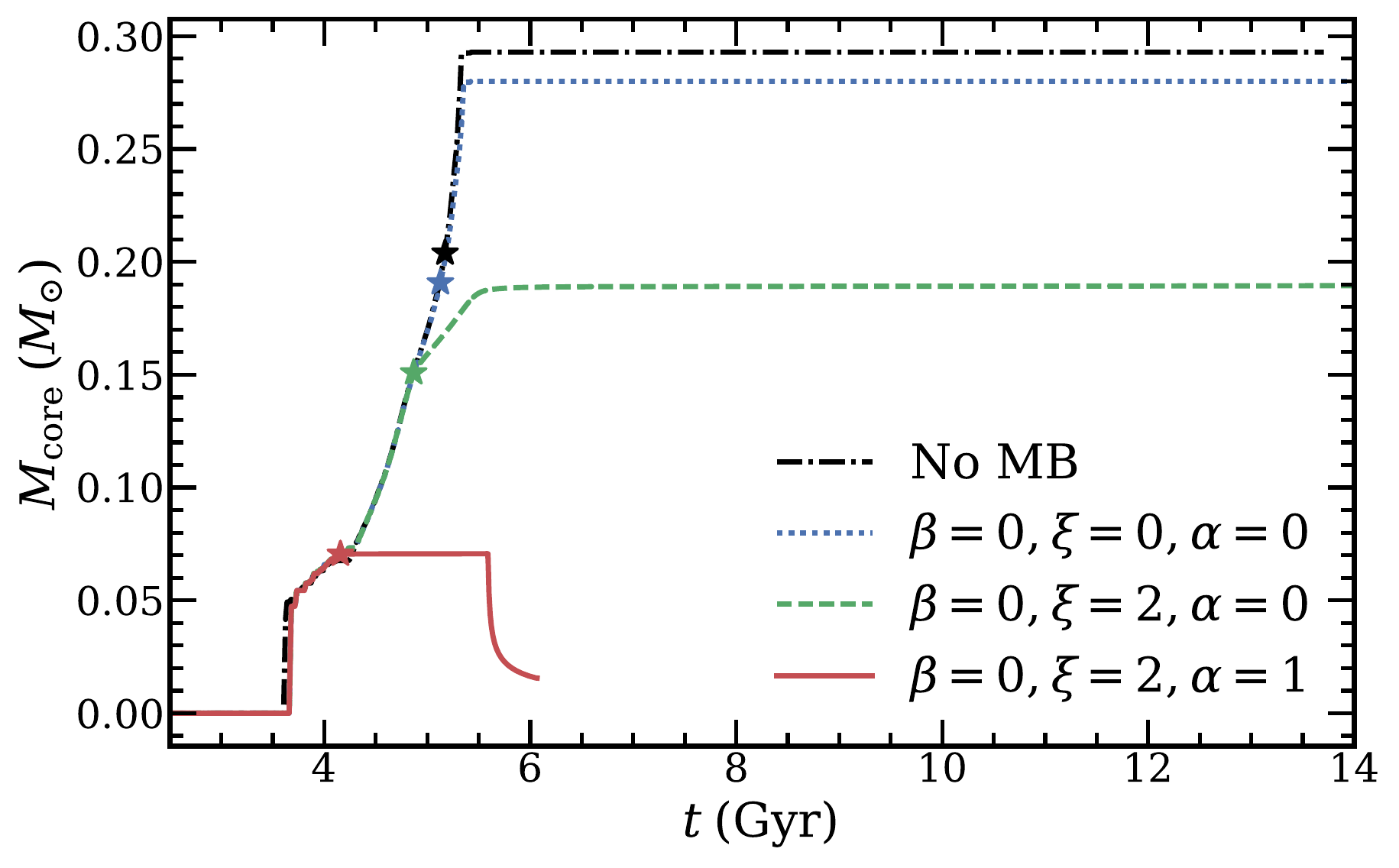}
\caption{Influence of different choices of free parameters in the equation of MB on binary evolution. In this plot, we also 
present the evolution of binaries without magnetic braking (black dash-dotted lines). 
The initial parameters of the binary systems are: $M_{\rm NS} = 1.30\;M_\odot$, $M_2 = 1.25\;M_\odot$, 
and $P_{\rm orb} = 5.0\;{\rm d}$.
The upper panel is the HR diagram. The middle and lower panels show the evolution of orbital period and He core mass as a function of time, respectively. 
The stars in the lower panel indicate the onset of mass-transfer phases. Different lines in the plot represent different combinations of MB parameters.
}
\label{fig:inf_diff_mb}
\end{figure}

\begin{figure}
\centering
\includegraphics[width=\columnwidth]{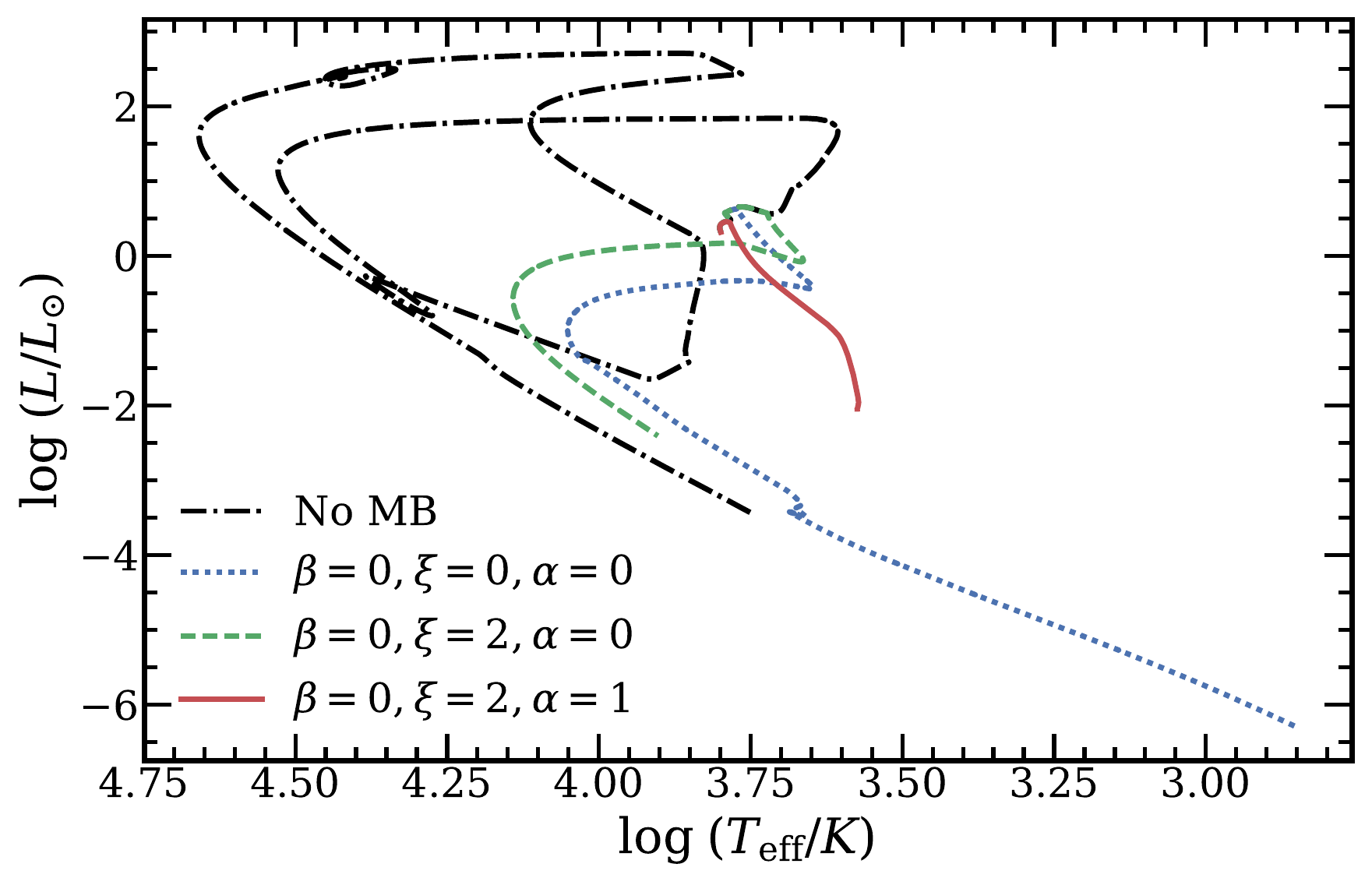}
\includegraphics[width=\columnwidth]{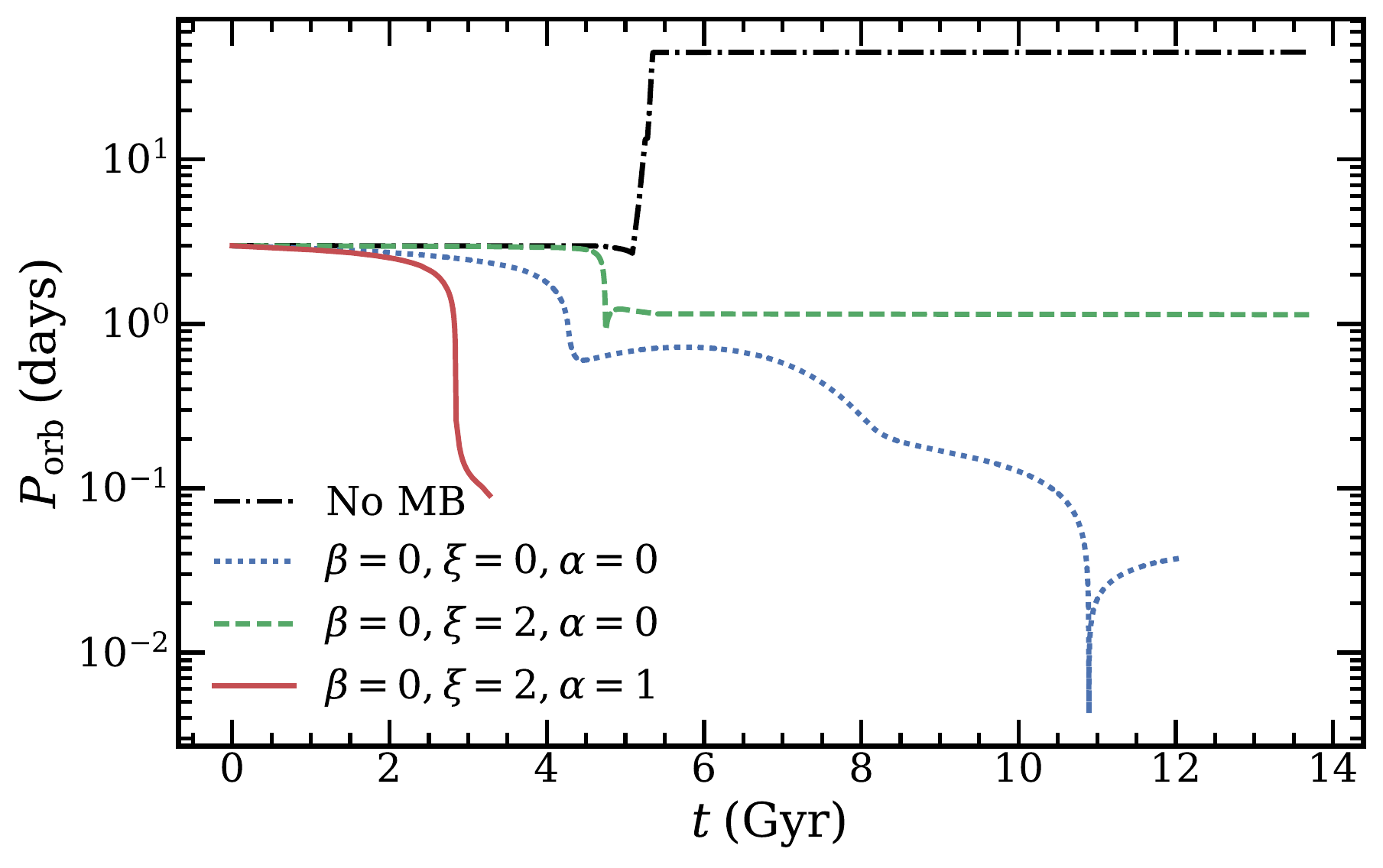}
\includegraphics[width=\columnwidth]{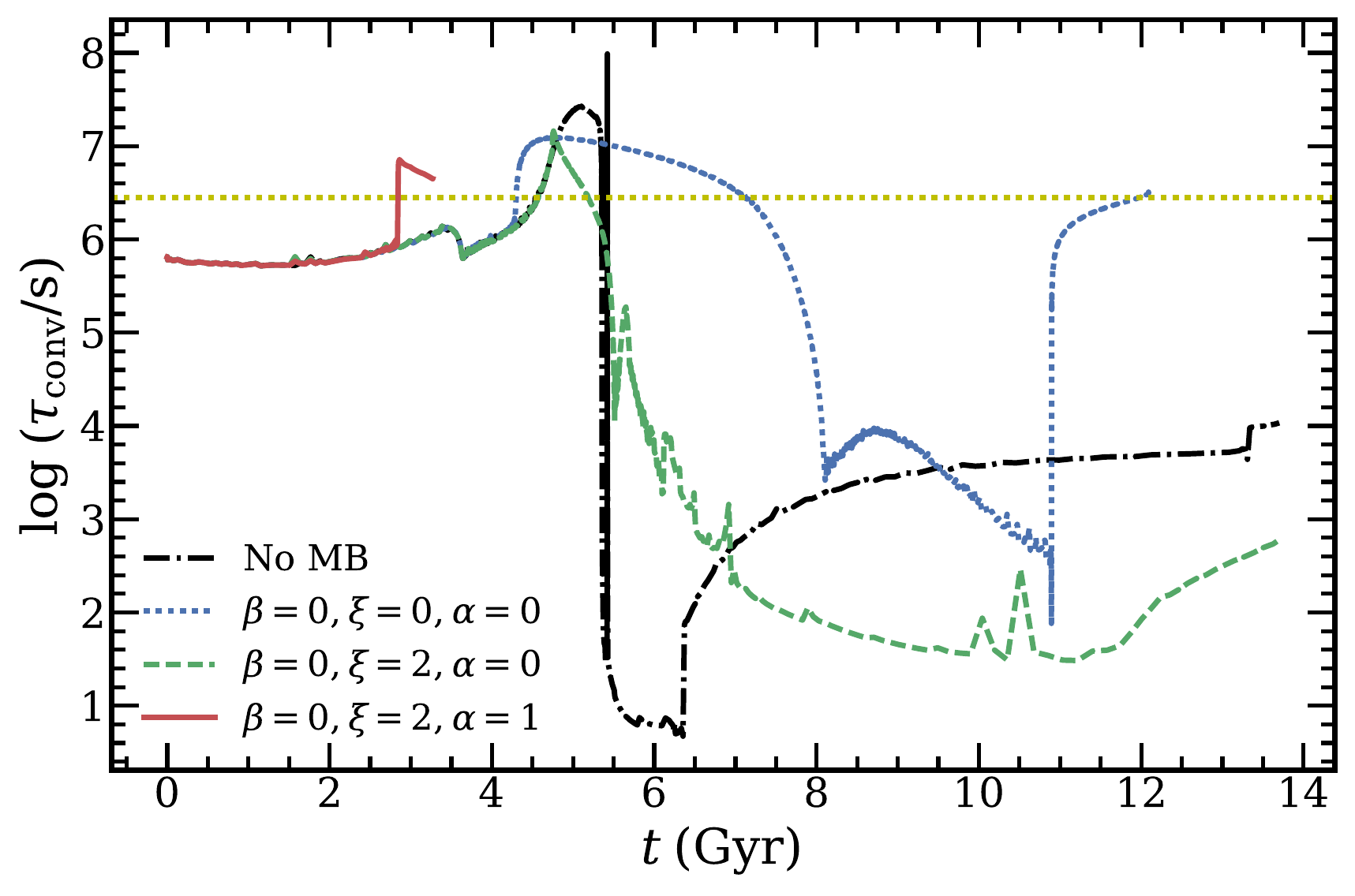}
\caption{Similar to Fig. ~\ref{fig:inf_diff_mb}, but the initial orbital period of the binaries is $P_{\rm orb} = 2.990\;{\rm d}$. The upper panel is the HR diagram. The middle panel shows the evolution of orbital period as a function of time. The lower panel shows the evolution of turnover time of convective eddies (see equation~\ref{eq:mb_vih}). The yellow dotted line in the lower panel represents the turnover time of the Sun. 
}
\label{fig:inf_diff_mb_sec}
\end{figure}

In Figs.~\ref{fig:inf_diff_mb} and ~\ref{fig:inf_diff_mb_sec},
we show the evolutionary tracks of binaries with different prescriptions of MB (no MB, MB1, MB2 and MB3). 
The initial NS and donor star masses in these two plots are the same as in Figs.~\ref{fig:bin_ev_ex}--\ref{fig:ele_bin_ex}, i.e. $M_{\rm NS} = 1.30\;M_\odot$ and $M_2 = 1.25\;M_\odot$.
The initial orbital periods are $P_{\rm orb} = 5.0\;{\rm d}$ in Fig.~\ref{fig:inf_diff_mb} and $P_{\rm orb} = 2.990\;{\rm d}$ in Fig.~\ref{fig:inf_diff_mb_sec}. Here we do not show tracks with $\beta \ne 0$ (e.g. MB4 with $(\beta,\,\xi,\,\alpha) = \,(2,\,4,\,1)$) since the computations for these evolutionary tracks do not converge. The main reason for this is the large values of $|\dot{M}_2|$ (see also \citet{vih19}).
In the prescription of MB including the scaling of a stellar wind (e.g. $\alpha =1$ in MB3), the orbital angular momentum loss becomes stronger as $|\dot{M}_2| > |\dot{M}_\odot|$ for more evolved donors. 
Thus with stronger orbital angular momentum loss due to MB, the binary orbital period decreases faster and the secondary star initiates RLO at an earlier evolutionary phase. This leads to different evolutionary fates of the systems depending on the MB prescription. 
However, the influence of the scaling of convective turnover time is complicated, as we will now discuss.

In the lower panel of Fig.~\ref{fig:inf_diff_mb_sec}, we present the evolution of the turnover time as a function of time for a donor star with $M_2=1.25\;M_\odot$. From this plot, we find that the turnover time is smaller than the turnover time of the Sun at early times ($t<4\;{\rm Gyr}$) and becomes larger at later times. Consequently, the MB will be weaker at early times and stronger at later times compared with the MB without including the scaling of the turnover time (i.e. $\xi = 0$ as for the standard MB prescription, MB1). This explains why the system with $\beta = \alpha =0$ and $\xi = 2$ (MB2, green dashed line) starts mass transfer earlier in Fig.~\ref{fig:inf_diff_mb} and later in Fig.~\ref{fig:inf_diff_mb_sec} compared with the MB1 model with $\alpha = \xi = \beta = 0$ (blue dotted line). 

\begin{figure}
    \centering
    \includegraphics[width=\columnwidth]{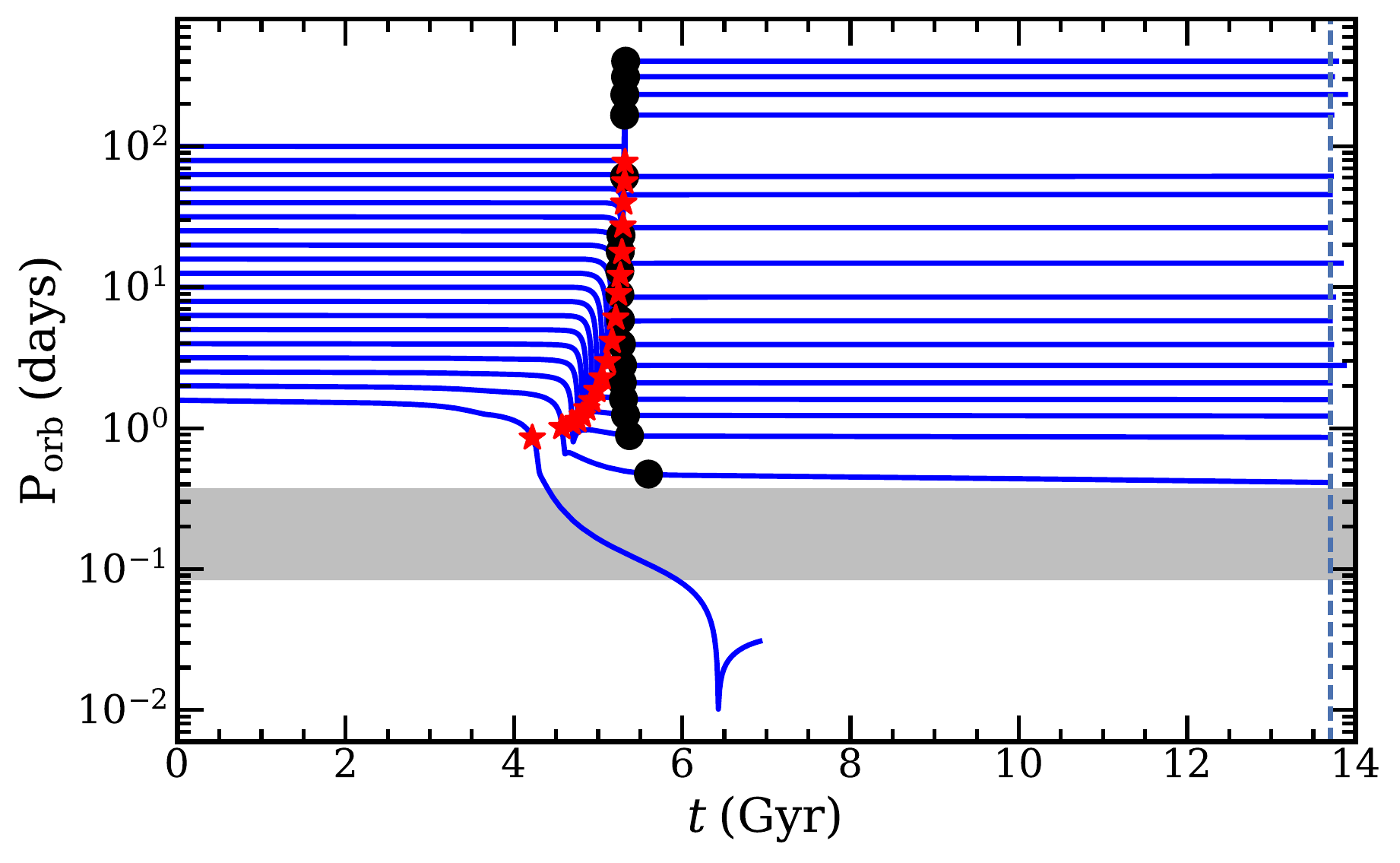}
    \caption{Evolutionary tracks calculated for binaries similar to those in Fig.~\ref{fig:t_porb_diff_porb}, but calculated with the MB2 prescription (see equation~\ref{eq:MB-laws}). Applying this weaker MB prescription (compared to MB3 in Fig.~\ref{fig:t_porb_diff_porb}) enables us to reproduce the wide-orbit BMSPs --- here shown up to $P_{\rm orb}\simeq 400\;{\rm days}$.}
    \label{fig:porb_evo_conv_boost}
\end{figure}

In Fig.~\ref{fig:porb_evo_conv_boost}, we present the evolution of orbital periods for models with $\beta = \alpha = 0$ and $\xi = 2$ (MB2). 
In these models, there are no problems producing divergent systems which detach and reproduce the observed wide-orbits BMSPs. This outcome is different from models with $\beta = 0$, $\xi = 2$ and $\alpha = 1$ (MB3, see Fig.~\ref{fig:t_porb_diff_porb}) which we found could not reproduce these wide-orbit BMSPs.

\begin{figure}
\centering
\includegraphics[width=\columnwidth]{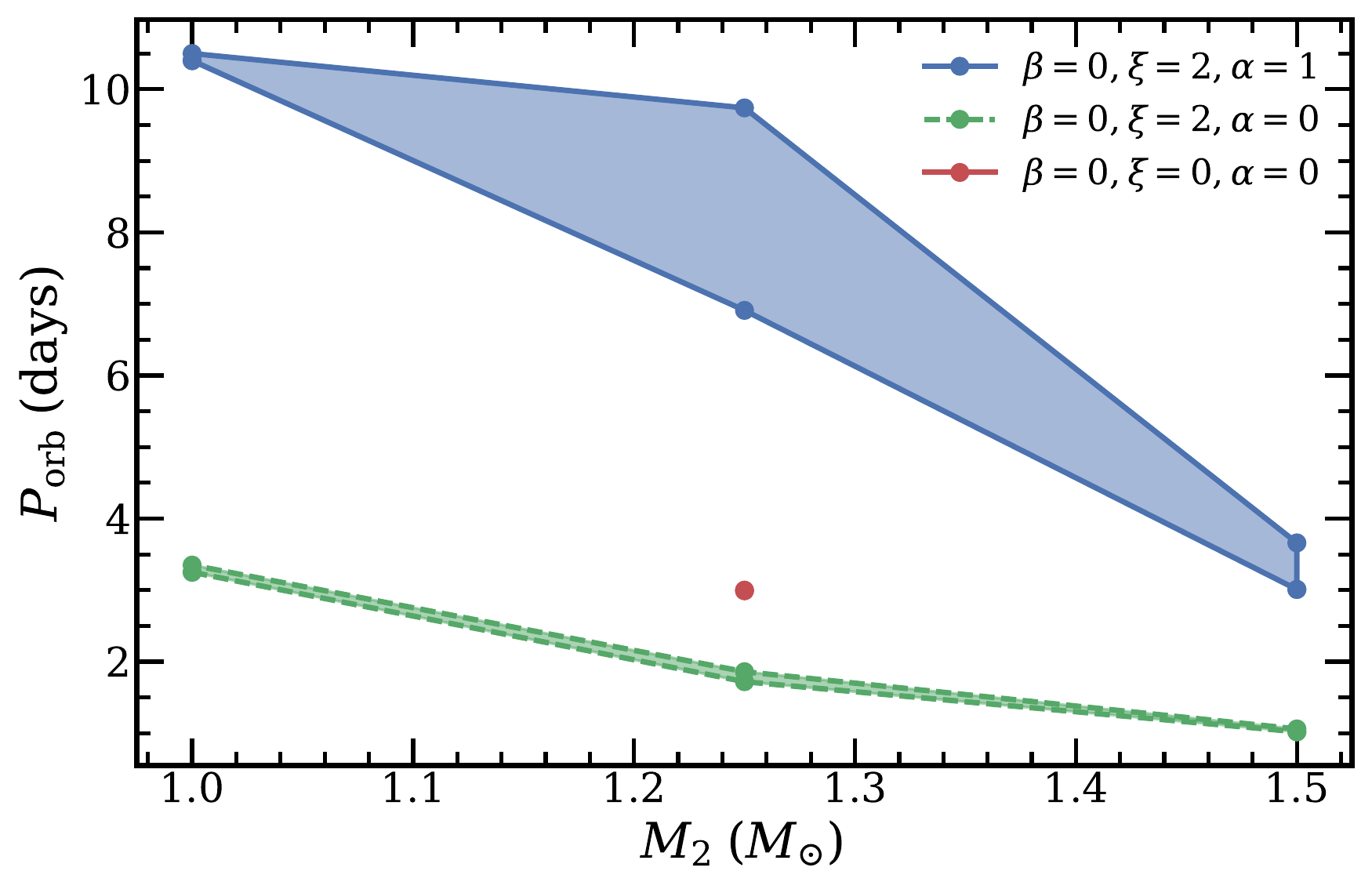}
\caption{Influence of different choices of free parameters in the prescription of MB (equation~\ref{eq:MB-laws}) on the initial parameter            space of binaries which evolve into detached NS+He~WDs and then UCXBs: MB3 (blue shaded area); MB2 (green shaded area), and MB1          (tiny red area near $P_{\rm orb}\simeq 3.0\;{\rm d}$ and $M_2=1.25\;M_\odot$). In all cases, we assumed an initial NS mass of $M_{\rm         NS}=1.30\;M_\odot$. 
        If we assume that no MB is operating, there are no solutions.
}
\label{fig:ini_par_diff_mb}
\end{figure}

In order to understand the influence of MB on the initial parameter space of binaries which can evolve into LMXBs, detached NS+He~WDs and then UCXBs, 
we also computed two grids of binaries with different values of $\alpha$, $\beta$ and $\xi$. Similar to our investigation in Section~\ref{subsec:para_spa}, the step of donor mass in the grid is $\Delta{M_2} = 0.25\;M_{\odot}$. The initial NS mass is kept fixed at $M_{\rm NS} = 1.30\;M_\odot$.
Our results are shown in Fig.~\ref{fig:ini_par_diff_mb}, which presents the initial parameter space of binaries which can evolve all the way to UCXBs, for models with different prescriptions of MB. For models with $\alpha = \beta = \xi = 0$ (MB1), we only find systems with $M_2 = 1.25\;M_{\odot}$ and initial orbital period, $P_{\rm orb} = 2.985 - 2.990\;{\rm d}$ can evolve into detached NS+He~WDs and then UCXBs. 
This narrow range of solutions is in analogy with the findings of \citet{itl14}; and see also \citet{tau18}.

In addition, we also checked the parameter space if the MB is turned completely off, and we found in that case that there are no systems which can evolve into detached NS+He~WDs (and UCXBs). 
We confirm in agreement with \citet{vvp05,vvp05b,itl14,ri19} that MB has an important impact on the formation of UCXBs.
Compared to models with relatively weak prescriptions of MB (MB1 and MB2), we find that the initial parameter space becomes larger if the MB is stronger (MB3). However, we have already concluded that the MB3 model cannot reproduce the observed wide-orbit BMSPs. Therefore, we disregard this model as a viable solution to produce UCXBs and GW bright sources.

\subsubsection{Perspectives for a universal MB prescription}\label{subsubsec:universal_MB}
We concluded earlier (Section~\ref{subsec:bin_ex}) that although the MB3 prescription relieves the fine-tuning problem in producing UCXBs, it introduces a new problem since, according to our computations in this study, it cannot explain the formation of the observed radio BMSPs in wide orbits. Hence, the MB3 prescription in its present form fails as a viable alternative candidate for a universal MB prescription, despite its advantages in explaining the observed persistent LMXBs \citep{vih19}. 

Finally, we note that any universal MB prescription must also be able to recover the characteristics of other low-mass binary systems, such as cataclysmic variables (CVs) and their orbital period gap \citep{kbp11}, polars \citep{bsp+20}, and post-CE binaries \citep{sgr+10}. 
At present, it seems that for some classes of binary systems a less efficient MB prescription is needed, while for others a more efficient MB prescription can better describe the observations (cf. the four above references).

\subsection{Influence of stellar rotation}
    
\begin{figure}
    \centering
    \includegraphics[width=\columnwidth]{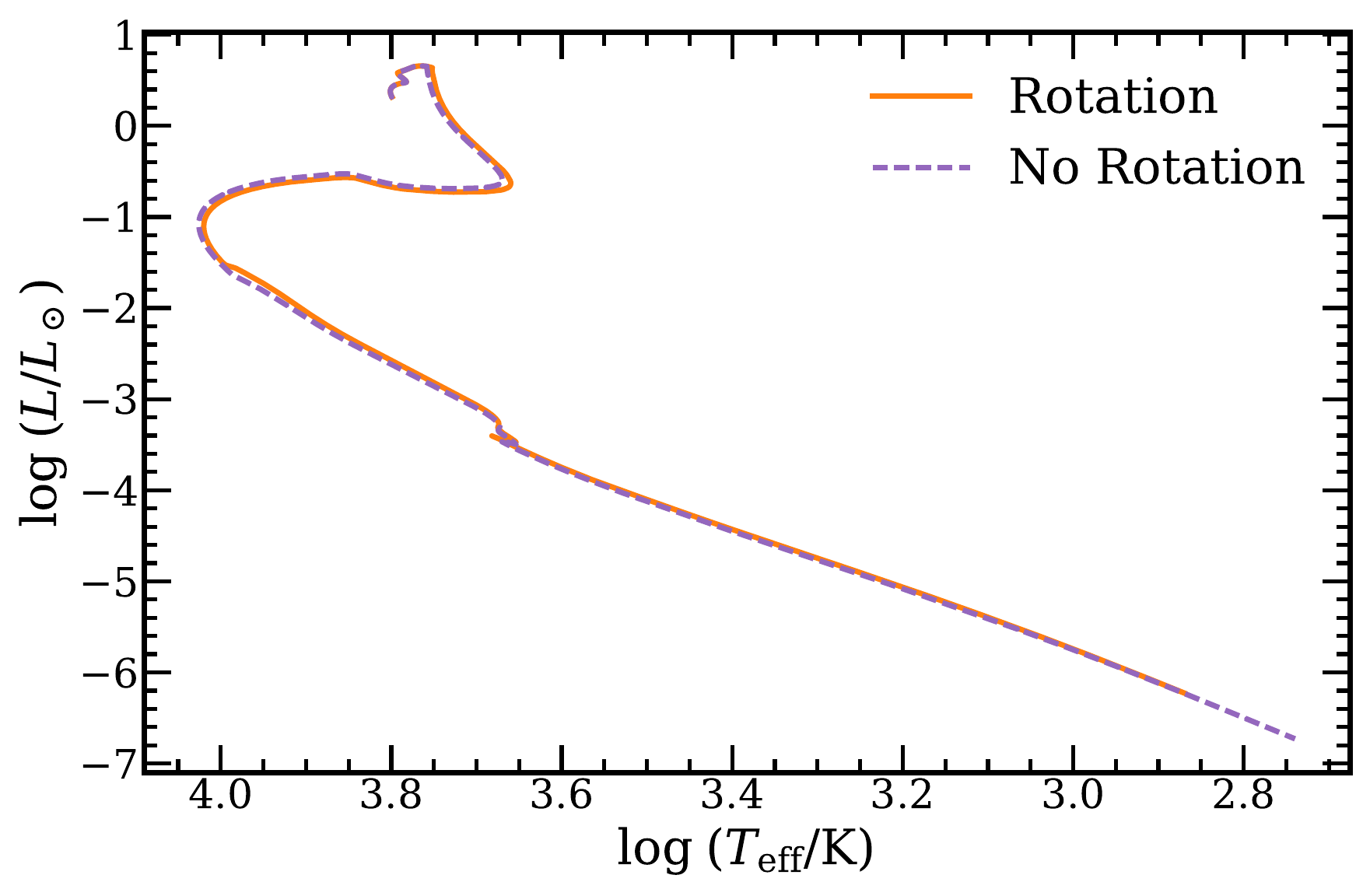}
    \includegraphics[width=\columnwidth]{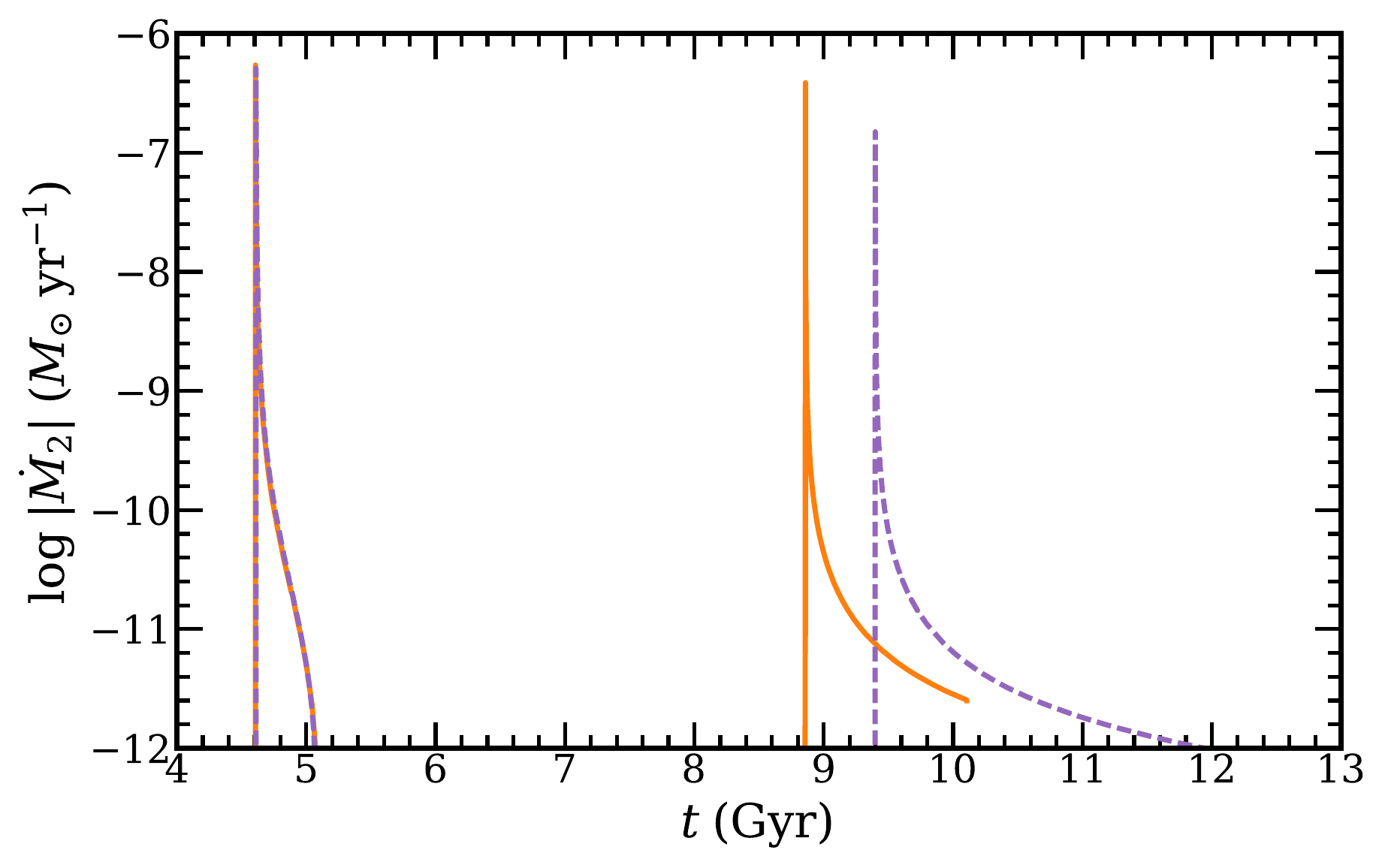}
    \includegraphics[width=\columnwidth]{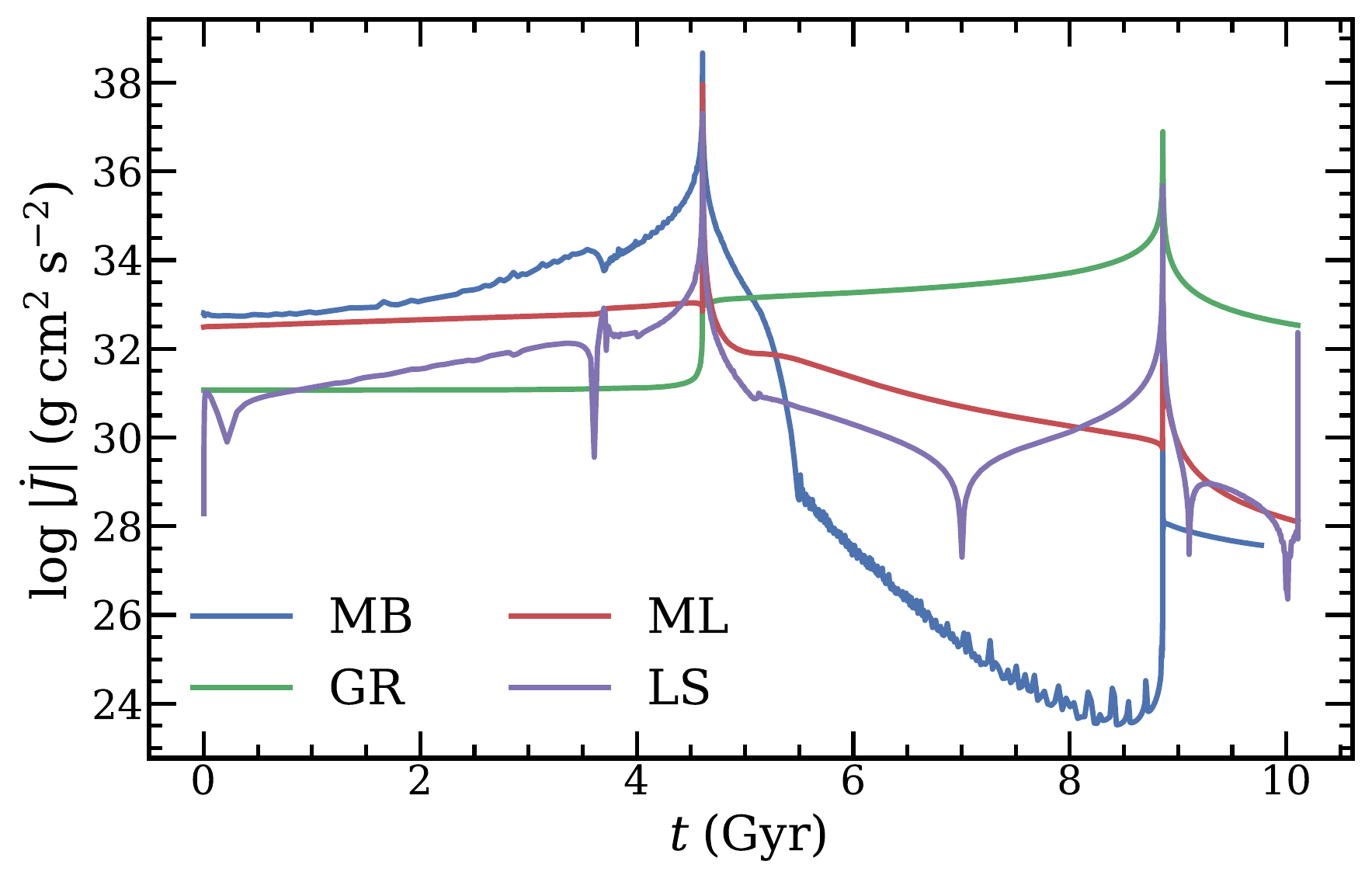}
    \caption{Comparison of binary evolution with and without stellar rotation. In these two examples, the initial binary parameters are $M_{\rm NS} = 1.30\;M_{\odot}$, $M_{\rm 2} = 1.25\;M_{\odot}$ and $P_{\rm orb} = 7.94\;$d. The upper panel shows the evolution of the donor stars in the HR~diagram, and the middle panel shows the evolution of mass-transfer rate as a function of time. In the upper and middle panels, the orange solid and purple dashed lines are for the model with and without rotation, respectively. The lower panel shows the evolution of orbital angular momentum loss due to GW radiation (GR), mass loss (ML), magnetic braking (MB), and other spin-orbit couplings (LS) for the model including stellar rotation.}
    \label{fig:com_rot}
\end{figure}

In order to understand the role played by stellar rotation in our simulations, we computed for comparison a model without considering the rotation of the donor star, which is shown in Fig.~\ref{fig:com_rot}. In these two examples, the initial binary parameters are $M_{\rm NS} = 1.30\;M_{\odot}$, $M_{\rm 2} = 1.25\;M_{\odot}$ and $P_{\rm orb} = 7.94\;$d. We found that the differences between the computations with and without stellar rotation are rather small. In the model including rotation, we see that the orbital angular momentum loss due to additional spin-orbit couplings is much smaller than the angular momentum loss due to MB\footnote{Notice, for MB to operate the star is assumed ad hoc to be rotating.} and GW radiation from the lower panel of Fig.~\ref{fig:com_rot}. In the model without stellar rotation, the orbital angular momentum loss due to additional spin-orbit couplings is zero. This explains why the differences between the model with and without stellar rotation are rather small. 

On the other hand, \citet{imt+16} found that rotation is important for the evolution of proto-WDs after the first (LMXB) mass-transfer phase, in particular for their surface chemical abundances and surface gravity (radial extent). Because of rotation, the radius of the WD at the onset of the second (UCXB) mass-transfer phase in the model including rotation is slightly larger than that in the model without rotation. This explains why the UCXB mass-transfer phase occurs earlier in the model including rotation. (The LMXB mass-transfer process at $t\sim 4.5\;{\rm Gyr}$ is seen to occur at the same time/rate in each case.)

\subsection{He~WD masses of detached NS+He~WD systems}
\citet{tau18} found that the He~WD masses are around $0.162\pm0.005\;M_\odot$ for the detached NS+ELM~WD systems that evolve into UCXBs within a Hubble time. Here, we also calculated the He~WD masses of detached NS+ELM~WD binaries for models with different prescriptions of MB.
For models with MB1, we find that the He~WD masses are $0.1604-0.1609\;M_\odot$. For models with MB2, the He~WD masses are, however, only $0.1447-0.1573\;M_\odot$, whereas for models with MB3 the He~WD masses are found in a wider interval of $0.1428-0.1635\;M_\odot$. Our results are consistent with \citet{tau18} who applied a standard MB prescription (MB1), but our calculations also demonstrate the possibility of somewhat smaller He~WD masses if applying a stronger MB prescription such as MB2 or MB3. The reason for this is that for a stronger MB, the donor stars are less evolved by the time they fill their Roche lobes and initiate the LMXB phases (Fig.~\ref{fig:inf_diff_mb}, lower panel).
Furthermore, in general, the He~core mass increases less for models with stronger MB during the mass-transfer phase.
On the other hand, stronger MB prescriptions capture binaries with even larger initial orbital periods and bring them into the UCXB band, thus resulting in a wider range of post-LMXB WD masses compared to the standard prescription MB1.

\subsection{Influence of the NS mass}
\begin{figure}
\centering
\includegraphics[width=\columnwidth]{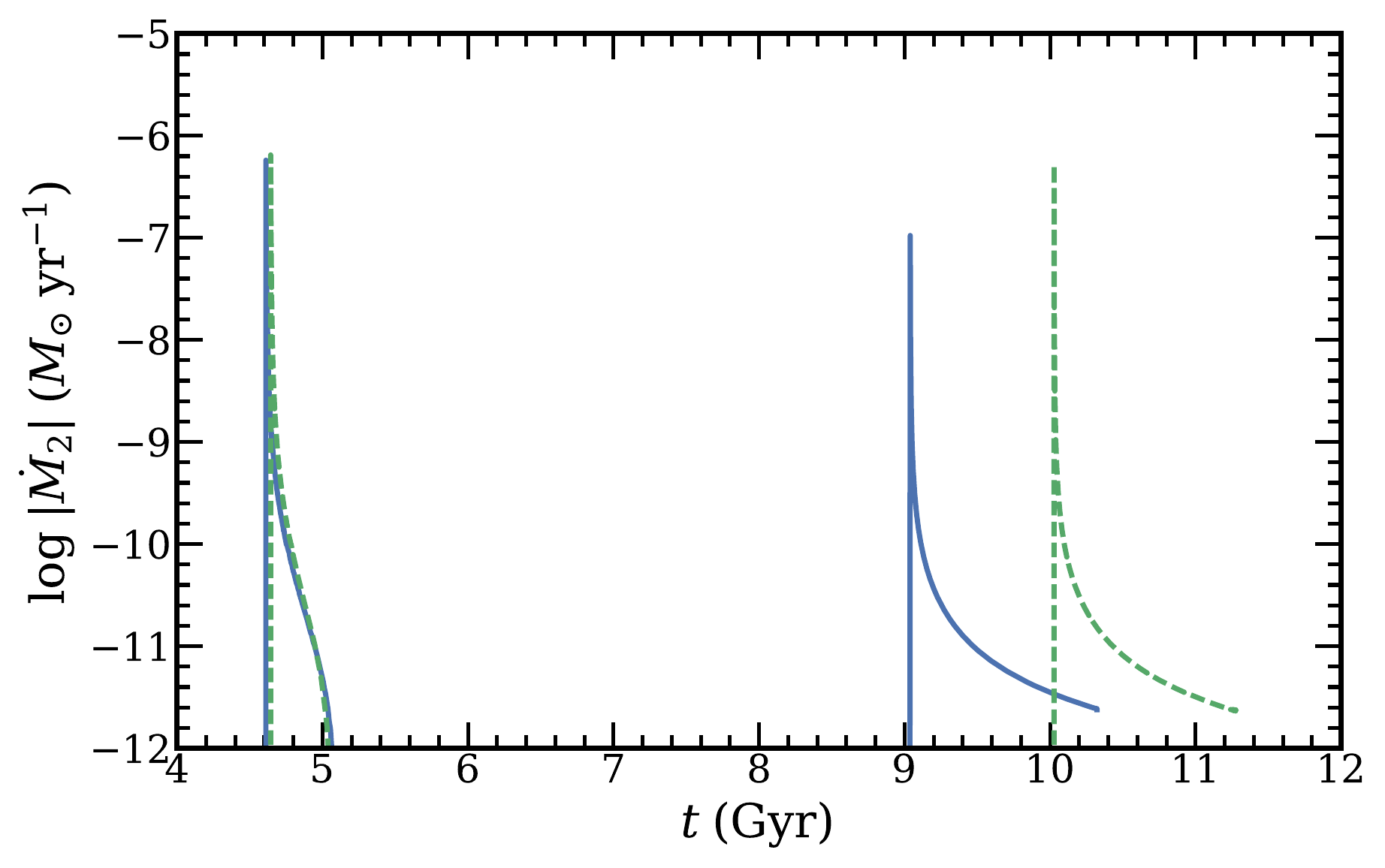}
\includegraphics[width=\columnwidth]{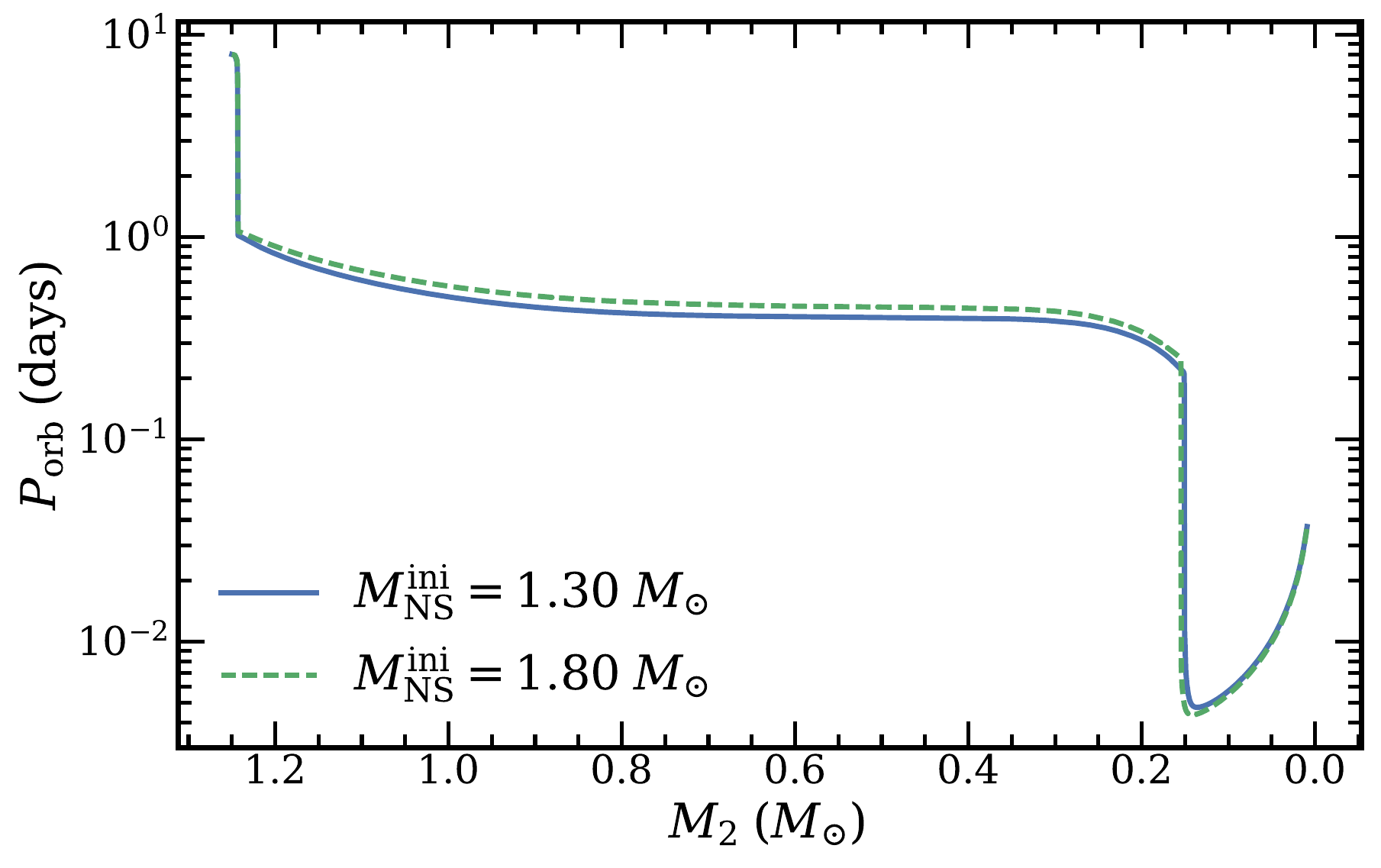}
\caption{Comparison of evolutionary tracks using MB3 of binary models with different initial NS masses. 
    The upper panel shows the evolution of mass-transfer rate as a function of time, and the lower panel shows orbital period as a function of decreasing donor mass.
	The initial donor star mass and orbital period are: $M_2 = 1.25\;M_\odot$ and $P_{\rm orb} = 8.0\;{\rm d}$.
	The blue solid and green dashed lines are for binaries with an initial NS mass of $1.30\;M_\odot$ and $1.80\;M_\odot$,
	respectively.}
\label{fig:com_diff_ns}
\end{figure}

In Fig.~\ref{fig:com_diff_ns}, we show the comparison of evolutionary tracks for binaries with different initial NS 
masses of $M_{\rm NS} = 1.30\;M_\odot$ and $M_{\rm NS} = 1.80\;M_\odot$. In both cases, the initial donor star mass is $M_2=1.25\;M_\odot$ and the initial orbital period is $P_{\rm orb}=8.0\;{\rm d}$. 
From this plot, we see that the difference between these two cases is rather small. In the model with $M_{\rm NS} = 1.80\;M_\odot$, it takes a longer time for the NS+WD system to become semi-detached. 
This can be understood as follows. The orbital period at the end of the first mass-transfer phase is larger for the system with $M_{\rm NS} = 1.80\;M_\odot$. The reason being that this system has a smaller mass ratio, $q\equiv M_2/M_{\rm NS}$ which causes a more efficient widening of the system during RLO \citep{tv06}. However, the He~WD masses at the end of the first mass-transfer phase are similar\footnote{One should here keep in mind the dependence of $q$ in Kepler's third law and in the relation for the Roche-lobe radius \citep{egg83}.}. Therefore, it takes longer time for the system with $M_{\rm NS} = 1.80\;M_\odot$ to become semi-detached due to GW radiation.
This explains why the upper boundary for the allowed initial parameter space in Fig.~\ref{fig:ini_par} is located at a smaller value of $P_{\rm orb}$ for models with $M_{\rm NS} = 1.80\;M_\odot$ compared to models with $M_{\rm NS} = 1.30\;M_\odot$. 

\subsection{$M_{\rm WD}-P_{\rm orb}$ relation}
\begin{figure}
\centering
\includegraphics[width=\columnwidth]{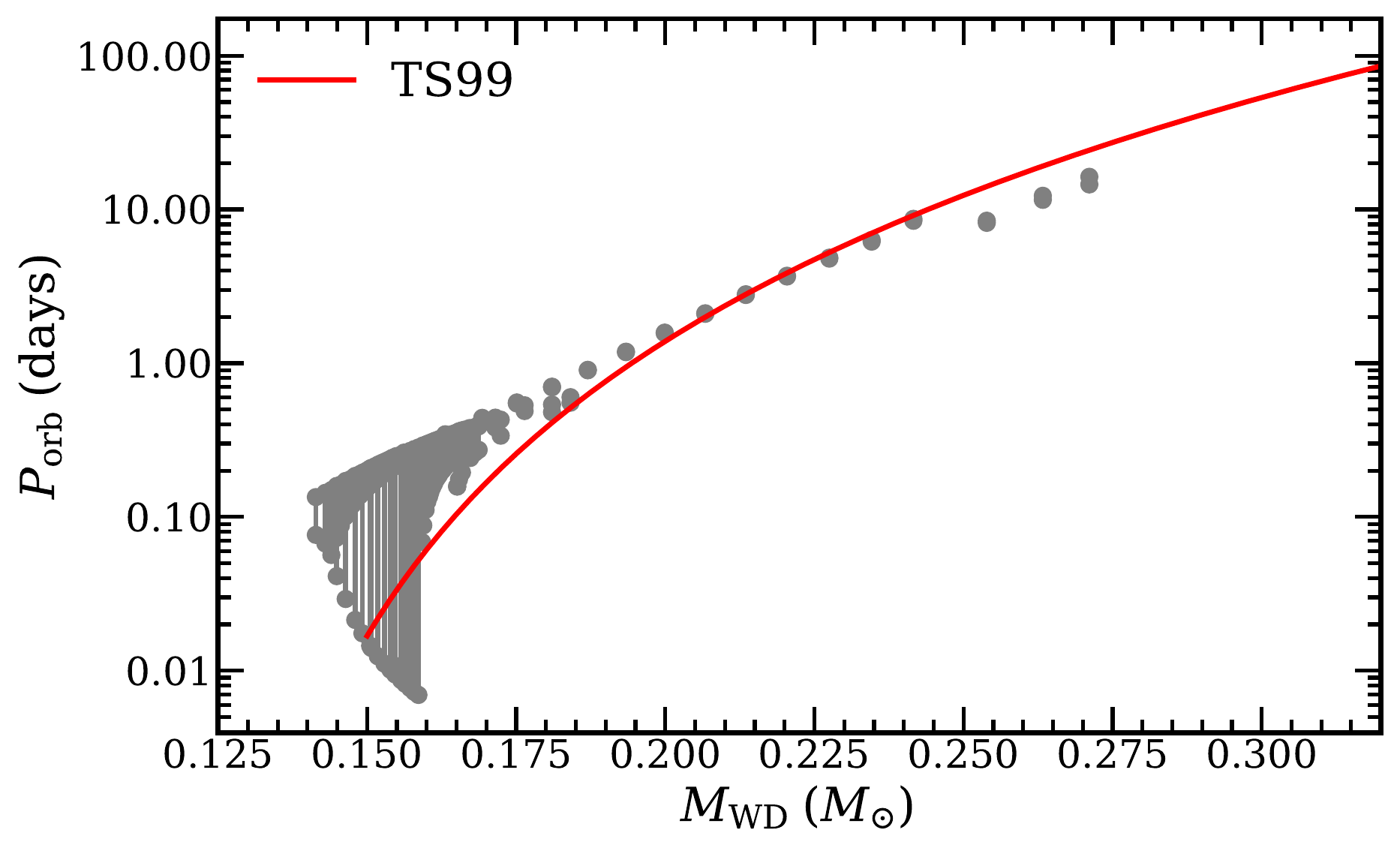}
    \caption{WD mass and orbital period relation for detached NS+He~WD systems, as calculated using MB3 and a wide range of initial donor star masses between $1.0-1.50\;M_\odot$, different initial values of $P_{\rm orb}$ and NS masses of $1.30\;M_\odot$. The solid red line is the relation for population~I stars from \citet{ts99}. For each NS+He~WD system, the maximum value of $P_{\rm orb}$ is the orbital period at the end of the LMXB phase. The minimum value of $P_{\rm orb}$ is the orbital period at the beginning of the UCXB phase; or after a Hubble time if it is still detached. }
\label{fig:mwd_lgp}
\end{figure}

Fig.~\ref{fig:mwd_lgp} shows the distribution of WD masses and orbital periods for the NS+He~WD systems calculated with the MB3 prescription, and for a range of different initial values of $P_{\rm orb}$, $M_2=1.0-1.50\;M_\odot$ and $M_{\rm NS}=1.30\;M_\odot$. For each detached system, the evolution from maximum and minimum orbital period that can be reached due to emission of GWs within a Hubble time are shown along a vertical line (i.e. the maximum value of $P_{\rm orb}$ is the orbital period when the He~WD is born; the minimum value of $P_{\rm orb}$ is the orbital period at the onset of RLO initiating the UCXB phase, or after a Hubble time if the system remained detached). 
The smallest orbital period among these systems before they initiate RLO and become UCXBs is around 10~min. Such small values in $P_{\rm orb}$ are obtained in systems where the ELM~WD had sufficient time to cool and contract to a size of $R_{\rm WD} = 0.038 \:R_\odot$.
It is also seen in this plot, that GW damping of the orbit is only significant for tight binaries, which correspond to systems with $M_{\rm WD}\la 0.16\;M_\odot$.  
For wider binaries with more massive He~WDs, their orbital periods remain almost fixed.
The red line shows the classical $M_{\rm WD}-P_{\rm orb}$ relation of \citet{ts99} for a chemical composition of Population~I stars. This relation was found to be valid for systems with $P_{\rm orb}>1\;{\rm d}$, which is confirmed by our calculations \citep[see also further evidence in][]{imt+16}. 

\subsection{Implication for the formation of double WDs with ELM~WD companions}                                                       
ELM~WDs are not only found in BMSPs, but also in other WD binaries with either A-star companions \citep[e.g.][]{rnls+15} or, in particular, in double WD systems \citep[e.g.][]{bkak10,bkak12,bkag+13,bgkk+16}. 
The formation of these ELM~WDs is also not well understood. 
\citet{lcch19} \cite[see also][]{sm18} have investigated the formation of ELM~WDs in double WD systems through the stable mass-transfer scenario and the CE scenario. 
They compared the theoretical $M_{\rm WD}-P_{\rm orb}$ relation with the observed data in their figs.~4 and 14, respectively, and found that 
the observed data is systematically below the theoretical relation. However, the observed systems all have $P_{\rm orb}\la 1\;{\rm d}$ and most of these systems are very likely to have been produced via CE evolution rather than stable RLO in CV-like systems --- especially the systems with WD masses above the typical ELM~WD limit of $\sim 0.20\;M_\odot$. Finally, among the remaining ELM~WD systems in tight binaries, many systems will have their orbits reduced significantly by GW damping. Therefore, it is not so surprising that many of the observed systems do not follow the theoretical $M_{\rm WD}-P_{\rm orb}$ relation which is based on stable RLO evolution.
In our next paper (in preparation), we will present our binary models for the formation of ELM~WDs in double WD systems.

\subsection{GW detection of UCXBs with space-based observatories}
\begin{figure*}
\centering
\includegraphics[width=2.0\columnwidth]{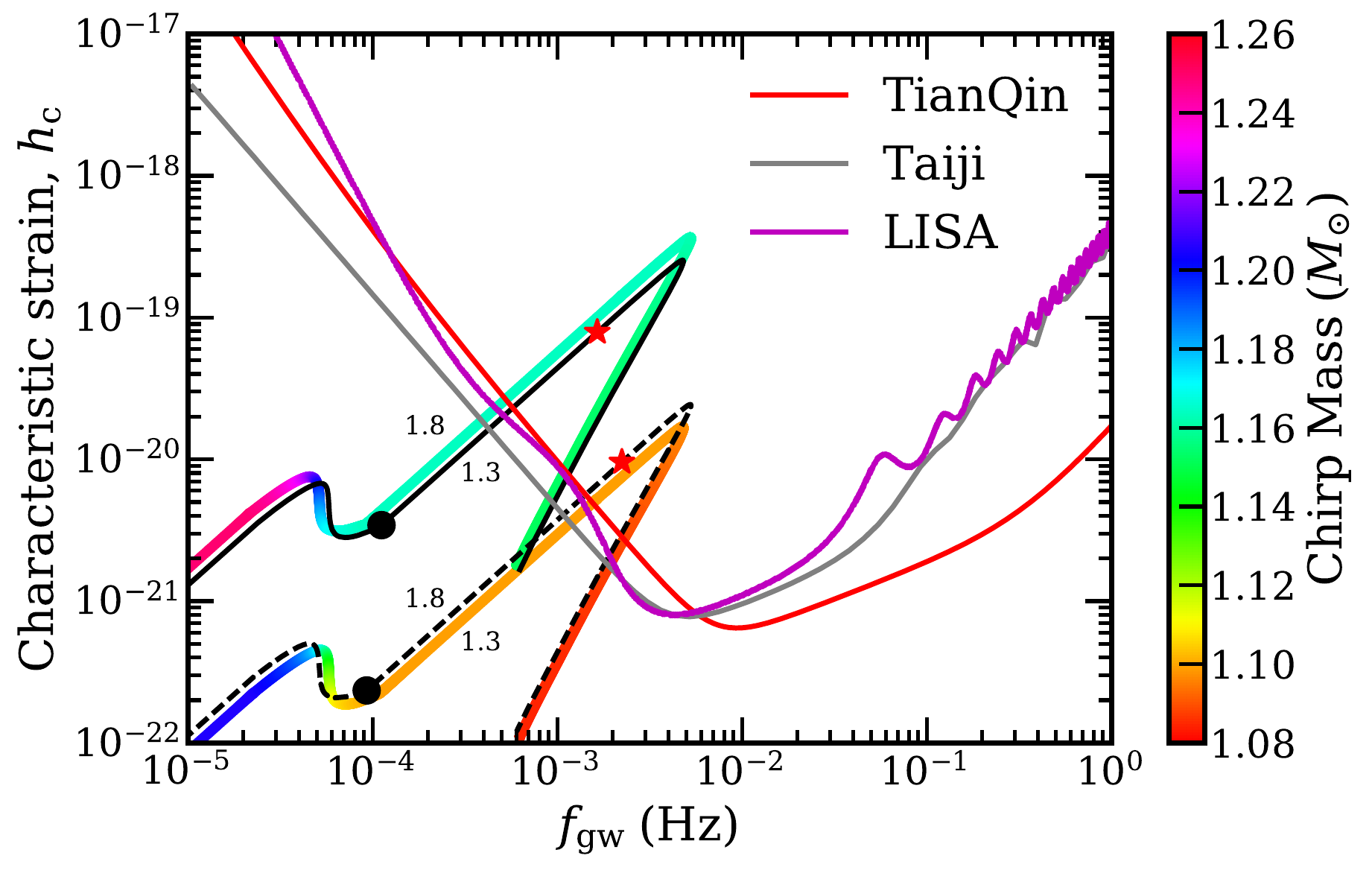}
\caption{Characteristic GW strain produced by UCXBs as a function of GW frequency. The black line, and also the thick coloured line below the black dashed line (marked ``1.3''), are for binaries with initial binary parameters of $M_{\rm NS} = 1.30\;M_\odot$, $M_2 = 1.25\;M_\odot$, $P_{\rm orb} = 7.94\;{\rm d}$. The black dashed line, and also the thick coloured line above the black solid line (marked ``1.8''), are for similar binaries with $M_{\rm NS} = 1.80\;M_\odot$. 
The upper and lower pairs of evolutionary tracks are for UCXBs at distances of $d = 1\;{\rm kpc}$ and $d = 15\;{\rm kpc}$, respectively. 
Their peak SNR are above 100 and 10, respectively.
The black circles represent the end of the first mass-transfer phase (LMXB) and the red stars indicate the onset of mass transfer in the UCXB phase. 
The colors on the two thick evolutionary tracks represent chirp masses (cf. vertical colour bar on the right).
The red and grey solid lines are the sensitive curves for TianQin and Taiji with an observing time, $T = 5\;{\rm yr}$, and 
the magenta line shows the sensitive curve for LISA with an observing time, $T = 4\;{\rm yr}$.
}
\label{fig:fre_str_tq}
\end{figure*}

\citet{tau18} demonstrated that the GW signal from Galactic UCXBs (and AM~CVn sources) are strong enough to be detected by LISA and studied their GW spectra as a function of time. Besides LISA, there are  
other space-based observatories going to be lunched in the next two decades, including TianQin and Taiji (see Section~\ref{sec:int}). 

In Fig.~\ref{fig:fre_str_tq}, we show the characteristic GW strain of an UCXB (including the detached pre-UCXB phase) as a function of GW frequency. 
The sensitivity curves of TianQin and Taiji with observing times, $T = 5\;{\rm yr}$, and LISA with an assumed observing time, $T = 4\;{\rm yr}$, are also plotted for comparison \citep{hhk+20}. 
The maximum signal-to-noise ratio (SNR) for such UCXBs at 1~kpc and 15~kpc are above 100 and 10, respectively.
This means that the far majority of all UCXBs in our Galaxy which are produced via stable RLO in LMXBs should be detected by TianQin, Taiji and LISA. Their pre-UCXB chirp masses are expected to be between $1.10-1.16\;M_\odot$ and their He~WD masses, after onset of the UCXB phase and for detection with LISA, are in the range $0.025-0.16\;M_\odot$ ($0.05 - 0.16\;M_\odot$), if these sources are located at 1~kpc (15~kpc).

\section{Summary and Conclusion}
\label{sec:con}
With the newly suggested prescription of MB from \citet{vih19}, we applied MESA to model the evolution of LMXBs. In particular, we studied the formation of NSs in tight orbits with ELM~He~WDs, i.e. BMSPs that later evolved into UCXBs as bright GW sources. 
We computed grids of binary evolution with different initial NS masses, donor star masses and orbital periods. 
Our main results are summarized as follows (listed in order by topic rather than importance).

\begin{enumerate}
    \item Our investigation of LMXB evolution with different MB prescriptions (MB1 to MB4, i.e. from standard and relatively weak MB to highly efficient MB) gave significantly different outcomes (cf. Figs.~\ref{fig:inf_diff_mb} and \ref{fig:inf_diff_mb_sec}). This result confirms the importance of determining the nature and the strength of MB for detailed calculations of LMXBs that evolve into BMSPs, UCXBs and GW sources. 
    
    \item Our evolutionary tracks calculated with the most efficient MB prescription of \citet{vih19}, MB4 with $(\beta,\,\xi,\,\alpha) = \,(2,\,4,\,1)$ did not cover a full grid of initial parameter values. However, those models that we calculated lead to numerical instabilities, as these models did not converge after onset of RLO due to very large values of $|\dot{M}_2|$ -- (possibly) leading to onset of a CE.
    
    \item If we adopt the relatively efficient MB3 prescription from \citet{vih19}, which includes MB scaling factors for the turnover timescale of convective eddies and wind mass loss, then, compared to models with standard MB prescription \citep[e.g.][]{rvj83}, the initial orbital period range of LMXBs which can evolve via detached NS+WD binaries into UCXBs becomes significantly wider (Fig.~\ref{fig:ini_par_diff_mb}) and thus helps to relieve the fine-tuning problem \citep{vvp05,vvp05b,itl14}. 
    
    \item In addition, applying the relatively efficient MB3 prescription from \citet{vih19} also results in a wider range of ELM~He~WD masses ($0.143-0.164\;M_\odot$) in detached NS+WD GW sources, compared to a more narrow range of WD masses expected from standard MB evolution. 

    \item In contradiction to the results of \citet{ri19}, our computed LMXB tracks with MB3 prescription are unable to produce wide-orbit BMSPs (Figs.~\ref{fig:t_porb_diff_porb} and \ref{fig:mwd_lgp}). 
    
    \item As a consequence of not being able to reproduce the many observed wide-orbit BMSPs using the MB3 prescription, we conclude that although the MB3 prescription relieves the fine-tuning problem in producing UCXBs, in its present form it fails as a viable alternative candidate for a universal MB prescription \citep[despite its advantages in explaining the observed persistent LMXBs][]{vih19}. Further investigations into alternative MB prescriptions are therefore needed. Any universal MB prescription must also be able to explain other low-mass binary systems, such as: CVs, polars and post-CE binaries (Section~\ref{subsubsec:universal_MB}).
    
    \item If the MB prescription scales with the convective turnover timescale as $(\tau_{\rm conv}/\tau_{\rm \odot,conv})^\xi$ (equation~\ref{eq:mb_vih}), we find that the MB will be weaker at relatively early evolutionary times and stronger at later (giant-star) stages (Fig.~\ref{fig:inf_diff_mb_sec}) compared with MB prescription without including the scaling of the turnover time (MB1: $\xi =0$).
 
    \item The GW signal of basically all (pre-) UCXBs in our Galaxy will be detected by space-based GW antennas, such as LISA, TianQin and Taiji (Fig.~\ref{fig:fre_str_tq}). We find that differences in NS masses will lead to slight differences in the measured chirp masses of pre-UCXBs between $1.10-1.16\;M_\odot$. The upper limit of the He~WD masses for UCXB GW sources produced via stable RLO in LMXBs are $\sim 0.16\;M_\odot$, as also found by \citet{tau18}.
\end{enumerate}


\section*{Acknowledgements}
We would like to thank the referee for useful comments on stellar rotation, which helped improving the paper. We thank Alina Istrate and Zhenwei Li very much for helpful advice and discussions.
H.-L.C. gratefully acknowledges support and hospitality from Aarhus University. This work is partially 
supported by the National Natural Science Foundation of China (Grant No. 12090040,12090043,11521303,12073071,11873016,11733008), the Yunnan Province (Grant No. 202001AT070058), 
and the CAS light of West China Program, Youth Innovation Promotion Association of Chinese Academy
of Sciences (Grant no. 2018076). H.-L.C. acknowledges the computing time granted by the Yunnan Observatories
and provided on the facilities at the Yunnan Observatories Supercomputing Platform. 
T.M.T. acknowledges an AIAS-COFUND Senior Fellowship funded by the European Union's Horizon 2020 Research and Innovation Programme (grant no.~754513) and the Aarhus University Research Foundation.
We are grateful to the \textsc{mesa} council for the \textsc{mesa} instrument papers and website.

\section*{DATA AVAILABILITY STATEMENT}

The data underlying this article will be shared on reasonable request
to the corresponding author.


\bibliographystyle{mnras}








\bsp	
\label{lastpage}
\end{document}